\documentclass[11pt,a4paper]{article}
\pdfoutput=1

\usepackage{jheppub-nosort,amsthm}

\usepackage{amsmath,amssymb,amsthm,amscd}
\usepackage{tikz}
\tikzset{node distance=2cm, auto}
\usepackage{epsfig}
\usepackage{psfrag,comment}
\usepackage{graphicx}
\usepackage{caption}
\usepackage{subcaption}
\usepackage{mathrsfs}
\usepackage{xcolor}
\usepackage[normalem]{ulem}

\newcommand{\CB}{{\cal B}}

\newcommand{\CF}{{\cal F}}

\newcommand{\CN}{{\cal N}}

\newcommand{\CP}{{\cal P}}

\newcommand{\CR}{{\cal R}}
\newcommand{\CS}{{\cal S}}

\newcommand{\CZ}{{\cal Z}}

\def\BZ{{\mathbb Z}}
\def\BR{{\mathbb R}}
\def\BC{{\mathbb C}}
\def\BP{{\mathbb P}}

\newcommand{\be}{\begin{equation}}
\newcommand{\ee}{\end{equation}}
\newcommand{\ba}{\begin{aligned}}
\newcommand{\ea}{\end{aligned}}
\newcommand{\bea}{\begin{eqnarray}}
\newcommand{\eea}{\end{eqnarray}}
\newcommand{\bean}{\begin{eqnarray*}}
\newcommand{\eean}{\end{eqnarray*}}

\def\r{\right\rangle}

\def\1{\mathbf{1}}
\def\0{|\1\r}
\def\im{{\mathbb{I}}{\mathrm{m}}}
\def\re{{\mathbb{R}}{\mathrm{e}}}

\newcommand{\rme}{{\rm e}}
\newcommand{\rmi}{{\rm i}}
\newcommand{\rmd}{{\rm d}}

\def\XXint#1#2#3{{\setbox0=\hbox{$#1{#2#3}{\int}$}
     \vcenter{\hbox{$#2#3$}}\kern-.5\wd0}}

\definecolor{pert}{cmyk}{0.5, 0.39, 0., 0.3}
\definecolor{1inst}{cmyk}{0.5, 0., 1., 0.4}
\definecolor{2inst}{cmyk}{0., 0.255, 0.75, 0.1}
\definecolor{3inst}{cmyk}{0., 0.6643, 0.73, 0.27}

\makeatletter
\newsavebox\myboxA
\newsavebox\myboxB
\newlength\mylenA

\newcommand*\widebar[2][0.75]{%
    \sbox{\myboxA}{$\m@th#2$}%
    \setbox\myboxB\null
    \ht\myboxB=\ht\myboxA%
    \dp\myboxB=\dp\myboxA%
    \wd\myboxB=#1\wd\myboxA
    \sbox\myboxB{$\m@th\overline{\copy\myboxB}$}
    \setlength\mylenA{\the\wd\myboxA}
    \addtolength\mylenA{-\the\wd\myboxB}%
    \ifdim\wd\myboxB<\wd\myboxA%
       \rlap{\hskip 0.8\mylenA\usebox\myboxB}{\usebox\myboxA}%
    \else
        \hskip -0.5\mylenA\rlap{\usebox\myboxA}{\hskip 0.5\mylenA\usebox\myboxB}%
    \fi}
\makeatother

\newdimen\tableauside\tableauside=1.0ex
\newdimen\tableaurule\tableaurule=0.4pt
\newdimen\tableaustep
\def\phantomhrule#1{\hbox{\vbox to0pt{\hrule height\tableaurule width#1\vss}}}
\def\phantomvrule#1{\vbox{\hbox to0pt{\vrule width\tableaurule height#1\hss}}}
\def\sqr{\vbox{%
  \phantomhrule\tableaustep
  \hbox{\phantomvrule\tableaustep\kern\tableaustep\phantomvrule\tableaustep}%
  \hbox{\vbox{\phantomhrule\tableauside}\kern-\tableaurule}}}
\def\squares#1{\hbox{\count0=#1\noindent\loop\sqr
  \advance\count0 by-1 \ifnum\count0>0\repeat}}
\def\tableau#1{\vcenter{\offinterlineskip
  \tableaustep=\tableauside\advance\tableaustep by-\tableaurule
  \kern\normallineskip\hbox
    {\kern\normallineskip\vbox
      {\gettableau#1 0 }%
     \kern\normallineskip\kern\tableaurule}%
  \kern\normallineskip\kern\tableaurule}}
\def\gettableau#1{\ifnum#1=0\let\next=\null\else
\squares{#1}\let\next=\gettableau\fi\next}
\preprint{
{\small{\textsf{YITP-SB-14-55}}}
}

\title{Finite N from Resurgent Large N}

\author[a]{Ricardo~Couso-Santamar\'\i a,}
\affiliation[a]{CAMGSD, Departamento de Matem\'atica, Instituto Superior T\'ecnico, Universidade de Lisboa,\\ Av. Rovisco Pais 1, 1049-001 Lisboa, Portugal\\}
\emailAdd{santamaria@math.tecnico.ulisboa.pt}

\author[a]{Ricardo~Schiappa,}
\emailAdd{schiappa@math.tecnico.ulisboa.pt}

\author[b]{Ricardo~Vaz\,}
\affiliation[b]{C.N.~Yang Institute for Theoretical Physics, Stony Brook University,\\ 
Stony Brook, NY 11794-3840, United States\\}
\emailAdd{ricardo.vaz@stonybrook.edu}

\abstract{
Due to instanton effects, gauge-theoretic large $N$ expansions yield asymptotic series, in powers of $1/N^2$. The present work shows how to generically make such expansions meaningful via their completion into resurgent transseries, encoding both perturbative and nonperturbative data. Large $N$ resurgent transseries compute gauge-theoretic finite $N$ results nonperturbatively (no matter how small $N$ is). Explicit calculations are carried out within the gauge theory prototypical example of the quartic matrix model. Due to integrability in the matrix model, it is possible to analytically compute (fixed integer) finite $N$ results. At the same time, the large $N$ resurgent transseries for the free energy of this model was recently constructed. Together, it is shown how the resummation of the large $N$ resurgent transseries matches the analytical finite $N$ results up to remarkable numerical accuracy. Due to lack of Borel summability, Stokes phenomena has to be carefully taken into account, implying that instantons play a dominant role in describing the finite $N$ physics. The final resurgence results can be analytically continued, defining gauge theory for any complex value of $N$.
}


\begin{document}

\maketitle



\allowdisplaybreaks

\section{Introduction and Discussion}\label{sec:intro}

The large $N$ expansion of nonabelian gauge theories was first uncovered four decades ago \cite{th74} (see, \textit{e.g.}, \cite{mz03} for a modern review). On the one hand, this expansion is topologically organized as a genus expansion which has led to remarkable developments in large $N$ duality, \textit{e.g.}, from supersymmetric gauge theories \cite{m97} to broad classes of matrix models \cite{dv02}. On the other hand, this expansion may be regarded as an approximation to compute gauge-theoretic observables nonperturbatively (from the point-of-view of the gauge coupling) at low values of the rank of the gauge group\footnote{As originally emphasized by 't~Hooft \cite{th74}, it was then hoped that the large $N$ expansion, and in particular the planar limit, would lead to ``reasonable'' approximations in spite of the physical interest in smaller values of $N$.}. While the first spin-off has been extremely successful over the years, the second has not received much attention beyond the planar approximation and its first few (mainly perturbative) corrections. It is our goal in this paper to introduce and develop a rather general method to extract (eventually \textit{exact}) finite $N$ results out of the large $N$ expansion.

The reason why at first sight the large $N$ expansion seems meaningless is because the resulting genus expansion is an asymptotic series, \textit{i.e.}, it has zero radius of convergence. Consider the free energy $F = \log Z$ of some nonabelian gauge theory, and denote by $t$ the 't~Hooft coupling (held fixed in the 't~Hooft limit). If one starts at fixed genus, say with a perturbative calculation of the planar limit $F_0 (t)$, this results in a convergent series in $t$ with non-zero radius of convergence \cite{knn77, th82}. But when one starts addressing the genus expansion of the free energy, one finds that its genus $g$ perturbative coefficients, $F_g (t)$, grow factorially fast as $F_g \sim (2g)!$. This results in a divergent asymptotic series which is not even Borel summable \cite{gp88, s90}. How can one make sense of such an expansion, in order to obtain finite $N$ results out of the large $N$ expansion?

For concreteness let us focus upon the free energy, whose large $N$ asymptotic expansion is
\be
\label{initial_perturbative}
F (N,t) \simeq \sum_{g=0}^{+\infty} N^{2-2g} F_g(t).
\ee
\noindent
One might hope to make sense out of this factorially divergent series via Borel resummation. In fact, the Borel transform removes the factorial growth from the perturbative coefficients as
\be
\label{borel_definition}
\CB [N^{2-2\alpha}] (s) = \frac{s^{2\alpha-3}}{\left( 2\alpha-3 \right)!},
\ee
\noindent
so that its (non-zero) radius of convergence is now dictated by the subleading exponential growth of the original perturbative coefficients. Upon analytic continuation of $\CB [F] (s)$ along the complex Borel $s$-plane, the Borel resummation follows whenever the ray of integration along the direction $\theta$ avoids the integrand's singularities. In this case one finds
\be
\label{resummation_definition}
\CS_\theta F (N,t) = \int_{0}^{\rme^{\rmi\theta} \infty} \rmd s\, \CB [F] (s)\, \rme^{-sN}.
\ee
\noindent
The important point is that even when the choice of direction $\theta$ is such that the above integration is possible, \textit{i.e.}, that one finds Borel summability along $\theta$, this expression need not match the exact result due to the occurrence of Stokes phenomena \cite{s64}. In fact, the factorial large-order behavior was already telling us that one is missing nonperturbative corrections, typically of the form $\sim \exp \left( -N \right)$, which are invisible to the perturbative expansion \eqref{initial_perturbative}. But in spite of that, these initially exponentially suppressed terms may not be neglected. In fact, upon variation of $\theta$ in \eqref{resummation_definition}, they may grow to become of order one or, eventually, exponentially enhanced with respect to \eqref{initial_perturbative}, and thus completely obliterate any sense in which the resummation \eqref{resummation_definition} may yield correct results. This is the essence of Stokes phenomena, implying that the perturbative series alone cannot properly define the free energy and it needs to be enlarged into a \textit{transseries}.

That Borel resummation, if allowed, cannot be the whole story when it comes to the genus expansion was recently verified numerically to great precision in \cite{gmz14} (following earlier work on the quantum mechanical quartic oscillator \cite{bpv78a, bpv78b}). In the example of the large $N$ ABJM gauge theory (but having also addressed other examples, such as topological strings in the local $\BP^1 \times \BP^1$ Calabi--Yau geometry), \cite{gmz14} showed numerically that even if this case satisfies the requirements of Borel summability, Borel resummation of the genus expansion does not agree with the exact value (computed via integrability), with the mismatch controlled by complex D2-brane instantons. As explained in the above paragraph, this is just explicitly verifying that, in general, Borel resummation is somewhat useless from a practical point-of-view due to Stokes phenomena: even if there are no obstructions to performing the inverse Borel transform in \eqref{resummation_definition}, and the asymptotic expansion is dubbed Borel summable, the resulting resummation will yield the correct result \textit{only if} Stokes phenomena has not yet occurred. Otherwise it will always miss nonperturbative contributions: further terms in the transseries may have already been turned on by the crossing of a Stokes lines and these are thus needed and crucial to obtain correct results.

Thus, building upon work initiated in \cite{m08} and as we shall discuss in the main body of the paper, finite $N$ results can only be obtained out of transseries; schematically of the form
\be
\label{initial_transseries}
F (N,t) = \sum_{g=0}^{+\infty} N^{2-2g} F^{(0)}_g(t) + \sigma\, \rme^{- N A(t)} \sum_{g=0}^{+\infty} \frac{1}{N^{g}}\, F^{(1)}_g(t) + \sigma^2\, \rme^{- 2 N A(t)} \sum_{g=0}^{+\infty} \frac{1}{N^{g}}\, F^{(2)}_g(t) + \cdots.
\ee
\noindent
Here $A(t)$ is the large $N$ instanton action\footnote{In those particular cases where the gauge theory is given by a matrix model, then the instanton action has a very physical description given in terms of eigenvalue tunneling; see \cite{d91, d92, msw07, msw08}.} and $F^{(n)}_g(t)$ the perturbative $g$-loop amplitudes around the nonperturbative $n$-instanton sector. Generically all these perturbative series are   asymptotic.  At large $N$, the exponential terms are non-analytic (thus the term ``transseries''), and the transseries parameter $\sigma$ encodes Stokes phenomena (we shall discuss this point carefully in the main body of the text, but see also \cite{as13} for general and very explicit formulae). Perhaps one of the most important properties of these transseries is that they are \textit{resurgent}. This means that, chosen any perturbative or multi-instanton sector, the coefficients of its associated asymptotic series, $F^{(n)}_g(t)$, have their large-order growth precisely dictated by the instanton action and by (the low-orders of) ``nearby'' coefficients $F^{(n')}_{g'}(t)$. Conversely, decoding the asymptotics of a chosen sequence $F^{(n)}_{g}(t)$ into its different power-series (analytic) and exponential (non-analytic) contributions in $\frac{1}{g}$, one will find the ``resurgence'' of all other sectors with $n' \neq n$, $g' \neq g$. See, \textit{e.g.}, \cite{asv11} for very explicit large-order resurgence formulae. All distinct components of the transseries \eqref{initial_transseries} are thus very tightly bound together by a stringent web of resurgence relations.

Some light introductions to resurgent transseries may be found in, \textit{e.g.}, \cite{cnp93, ss03, e0801} with more mathematical flavor, or in the first sections of \cite{asv11} with a flavor closer to the present work. Resurgent transseries are at the basis of \'Ecalle's work \cite{e81}, where by joining analytic and non-analytic building blocks they allow for representations (in some sense, for constructions and \textit{definitions}) of broad classes of functions, including functions with singularities and branch cuts. This is why in many cases where one might only have large $N$ expansions available, a large $N$ resurgent transseries may allow for a \textit{definition} of the finite $N$ theory and (considering arbitrarily large numbers of components in the transseries) for a calculation of \textit{exact} finite $N$ results.

While the initial gauge theory was only defined for \textit{integer} values of $N$, the resurgent transseries is constructing a function of $N$, eventually valid upon analytic continuation for \textit{arbitrary complex} values of $N$. While at first this might sound unexpected for nontrivial interacting gauge theories, it is actually familiar in free gauge theories such as the Gaussian or Penner matrix models (see, \textit{e.g.}, \cite{m04, ps09} for brief reviews). For instance, the partition function of the Gaussian matrix model (normalized against the volume of its $\text{U}(N)$ gauge group), while only defined for integer values of $N$ via a matrix integral, may still be computed as\footnote{The so-called string coupling $g_s$ appears as the overall coupling in front of the quadratic potential inside the matrix integral defining the partition function, and relates to the 't~Hooft coupling as $t=g_s N$.}
\be
\label{gaussian_Z}
Z_{\text{G}} (N) = \frac{g_s^{\frac{N^2}{2}}}{\left( 2\pi \right)^{\frac{N}{2}}}\, G_2 \left( N+1 \right),
\ee
\noindent
where $G_2 (z)$ is the Barnes $G$-function which is in fact an entire function on the complex plane (see, \textit{e.g.}, \cite{olbc10}). Finding nontrivial gauge-theoretic examples where nonperturbative completions allow for results defined at arbitrary values of $N$ is not as straightforward. But a class of very interesting such examples was recently addressed in \cite{cgm14}, within the context of ABJM gauge theory on the three-sphere. In their examples, supersymmetry and integrability ensure that all (genus) contributions to the partition functions actually \textit{truncate} to finite sums, allowing for the computation of closed-form expressions at arbitrary values of $N$. In particular, partition functions for these theories were also found to be entire functions on the complex $N$-plane.

One fascinating spin-off of the calculations in \cite{cgm14} is the possibility to study exact properties of the partition function of ABJM gauge theory as a function of $N$. As shown therein, the fact that all partition functions found in \cite{cgm14} were entire functions on the complex $N$-plane implies the non-existence of phase transitions in $N$; in particular the non-existence of a phase transition as one moves from (real) large $N$ to small $N$. From a dual holographic viewpoint \cite{m97} this implies there is really no drastic distinction between a ``quantum'' small-distance and a ``semiclassical'' large-distance phase, at least in this class of theories. Furthermore \cite{cgm14}, this is particularly striking given that if one just looks at the genus expansion by itself, the perturbative free energies do have branch cuts in the 't~Hooft coupling $t$, misleading us on the understanding of the nonperturbative physics. This phenomenon where the perturbative singularity structure is nonperturbatively smoothened was also found earlier in \cite{mmss04}, within the context of minimal string theory. Therein, certain observables such as the FZZT partition function, while having semiclassical branched domains, end up having entire analytical properties when addressed nonperturbatively. Interestingly enough, the mechanism by which this occurs is again based upon the aforementioned Stokes phenomena, whereby (initially) exponentially small contributions grow to take dominance and dramatically change the structure of the solutions. The large $N$ resurgent transseries methods we develop in this paper will allow for these types of analysis in broad classes of gauge theories, even those where neither supersymmetry nor integrability are present.

One final point to discuss is whether the resurgent transseries completion we propose, to non-integer (and complex) values of $N$, is the only possible such completion? In principle, gauge theory observables are only defined at positive integer values of $N$, and, clearly, many possible interpolating functions may exist through such a discrete set of points. It was already suggested in \cite{cgm14} that the familiar Gamma function serves as a good analogy. Certainly there are many other functions which solve the factorial interpolation problem, and which in fact behave much better (the Gamma function is meromorphic). One such example is the Hadamard Gamma function, which interpolates the factorials as an entire function, but many others exist; see \cite{l06} for an excellent review. What tells all these possible interpolating functions apart is, of course, whatever (differential or functional) equation defines them. In this regard, the usual Euler Gamma function is the \textit{only}\footnote{To be correct, uniqueness is only achieved after adding the technical requirement of logarithmic convexity \cite{l06}.} function which satisfies the continuous (functional) version of the factorial recursion,
\be
\label{euler_gamma_function}
\Gamma ( z+1 ) = z\, \Gamma ( z ).
\ee
\noindent
When translating to our current large $N$ problem, a rather similar situation still holds. As we shall see in the main text of this paper, the partition function at finite integer $N$ is computed by solving a discrete recursion relation (the so-called string equation \cite{biz80}, but we shall discuss this at greater length later), while the large $N$ resurgent transseries is obtained out of the \textit{continuous} (functional) version of this discrete string equation \cite{m08, asv11, sv13}. Furthermore, this solution is very special, in the sense that it is a \textit{resurgent} solution: all its perturbative and nonperturbative building blocks are very tightly bound together! Although we do not address this point in this work---and it should certainly be addressed in future research---it might be reasonable to expect that these two properties combined provide a \textit{unique} analytic continuation of gauge theoretic observables to complex $N$ (at least up to some choice of boundary conditions).

The problem of resummation of divergent asymptotic series in Physics has a long history; see, \textit{e.g.}, \cite{z81}. In order to apply it, and extend it via the use of resurgent transseries, to the problem of the large $N$ expansion one first needs data, on both perturbative and multi-instanton expansions. To the best of our knowledge, the gauge theories for which at present there is the largest amount of available large-$N$ resurgent-transseries data are matrix models (and their double-scaling limits); in particular matrix models with polynomial potentials \cite{m08, gikm10, kmr10, asv11, sv13}. This type of matrix models is a very reasonable prototype for generic gauge theories. We shall perform our explicit calculations in the example of the quartic matrix model \cite{asv11, sv13}, although we should stress that our methods are completely general within gauge theory.

This paper is organized as follows. We begin in section \ref{sec:exact} with a calculation of exact results in the quartic matrix model, at small (integer) values of $N$. This is done via standard methods using orthogonal polynomials, and several exact results are obtained for the partition function and the recursion coefficients. At fixed $N$, these quantities are still functions of the coupling, or, equivalently, of the 't~Hooft coupling, and we address their corresponding non-trivial monodromy properties. In section \ref{sec:resurgent} we then address the gauge-theoretic large $N$ resurgent solution to our example. While the resurgence data of the quartic matrix model is reviewed in appendix \ref{sec:appendix}, section \ref{sec:resurgent} shows how to make use of this data (or resurgence data from any other gauge theory) in order to address the resummation problem and produce finite $N$ results. Comparison with the analytical results is achieved up to remarkable numerical accuracy. Finally, section \ref{sec:resurgent_stokes} discusses Stokes phenomenon and how the full resurgent transseries is crucial to reproduce the analytical finite $N$ results across the complex $t$-plane, including the distinct non-trivial monodromy properties one found earlier for different integer values of $N$. We conclude with a discussion of how the large $N$ transseries may be analytically continued for any complex value of $N$, and what are the properties of the resulting partition function on the complex $N$-plane.

We believe that many interesting problems may be addressed in the near future using the methods we introduce and develop in this paper. For instance, one may consider gauge theories in higher dimensions. When considering supersymmetric gauge theories on compact manifolds where some observables localize \cite{p07}, we may fall back into the realm of matrix models. The resurgent analysis of localizable observables in some three and four dimensional gauge theories was initialized in \cite{ars14} and it would certainly be interesting to address finite $N$ calculations in that realm. If we move away from matrix models, the main difference will be that the (now diagrammatic) computation of perturbative and multi-instanton coefficients will be much more time consuming. Our uses of large $N$ resurgent transseries, however, should hold step by step. Perhaps a good route where to start would be to follow the recent resurgence calculations for the $\BC\BP^N$ model in \cite{du12a, du12b, cdu14} and address the corresponding large $N$ limit in that context.

Via large $N$ duality, another class of theories to explore are string theories. The simplest cases might be within topological string theory, whose resurgent analysis has also been steadily developed \cite{m06, msw07, m08, ps09, cesv13, cesv14} in recent years. In particular, the resurgence techniques introduced in \cite{cesv13} have allowed for the generation of large amounts of resurgence data concerning the local $\BP^2$ geometry \cite{cesv14}, and are easily extendable to generate equal amounts of data for other toric Calabi--Yau geometries such as local $\BP^1 \times \BP^1$. In this context, a very interesting comparison is actually now possible due to a recent proposal for obtaining nonperturbative results for topological strings in toric Calabi--Yau geometries \cite{ghm14} (following up on earlier work in \cite{hmmo13, km13}). This proposal allows for the calculation of topological string free energies at continuous values of $N$ for several local Calabi--Yau geometries; in particular it addresses the examples of local $\BP^2$ and local $\BP^1 \times \BP^1$. It would be extremely interesting to use the aforementioned resurgence data for these geometries \cite{cesv13, cesv14}, combined with the methods in the present paper, to investigate how accurate would be the match between both approaches, this time around at \textit{continuous} $N$.

\section{Exact Finite $N$ Results}\label{sec:exact}

In order to check both the validity and the resulting accuracy of any large $N$ transseries results, we need data to compare them against. As already mentioned in section \ref{sec:intro}, this will be done within the gauge-theoretic context of the quartic matrix model, where orthogonal polynomial techniques \cite{biz80} allow for straightforward calculations at small (integer) values of $N$. We refer the reader to \cite{m04} for an excellent overview of these methods.

Let us begin by recalling the definition of the partition function for a one-cut matrix model, with polynomial potential $V(z)$, already written in diagonal gauge with eigenvalues $z_i$ (as usual, the partition function was originally normalized against the volume of the gauge group $\text{U}(N)$). One has:
\be
\label{partitionfunctionV}
Z(N) = \frac{1}{N!} \int \prod_{i=1}^{N} \frac{\rmd z_i}{2\pi}\, \Delta^2(z)\, \rme^{-\frac{1}{g_s} \sum_{i=1}^{N} V(z_i)},
\ee
\noindent
where $\Delta^2(z) = \prod_{i < j} (z_i - z_j)^2$ is the Vandermonde determinant. In this expression we have chosen the standard convention (within the matrix model literature) of using the so-called string coupling, $g_s$, as the overall normalization, but this is of course just a notation choice and we could have likewise used a gauge-theoretic coupling $g_s \sim g^2_{\text{YM}}$. In any case, in the large $N$ 't~Hooft limit one has $t = g_s N$ fixed and we shall thus mostly work with the 't~Hooft coupling, $t$, and gauge group rank, $N$, in the following. Furthermore, we shall focus exclusively upon the quartic potential with quartic coupling\footnote{It is important to notice that there are in fact less independent couplings in the problem than apparent at first sight. Changing variables as $w=\frac{z}{\sqrt{g_s}}$, leads to
\be
Z(N) = \frac{g_s^{\frac{N^2}{2}}}{N!} \int \prod_{i=1}^{N} \frac{\rmd w_i}{2\pi}\, \Delta^2(w)\, \rme^{-\sum_{i=1}^{N} W(w_i)},
\ee
with a modified potential
\be
W (w) = \frac{1}{2} w^2 + \frac{1}{4!} \kappa w^4,
\ee
\noindent
where we have defined $\kappa \equiv -\lambda g_s$. It is then clear that there is a unique independent coupling in the matrix integral, although, as we said above, we shall mostly work with the 't~Hooft coupling $t$.} $\lambda$,
\be
\label{potentialV}
V(z) = \frac{1}{2} z^2 - \frac{1}{4!} \lambda z^4.
\ee

It is a straightforward task to introduce orthogonal polynomials $p_n(z) = z^n + \cdots$, satisfying
\be 
\int_{\BR} \rmd \mu(z)\, p_n (z)\, p_m (z) = h_n\, \delta_{nm}
\ee
\noindent
with respect to the measure $\rmd \mu (z) = \frac{\rmd z}{2\pi}\, \rme^{- \frac{1}{g_s} V(z)}$ inherited from the matrix potential. A simple calculation \cite{biz80, m04} then shows how the partition function may be written as a product of the above polynomial norms
\be 
\label{eq:Zhr}
Z(N) = \prod_{n=0}^{N-1} h_n = h_0^N\, \prod_{n=1}^{N} r_n^{N-n},
\ee
\noindent
or, equivalently, as a product of the recursion coefficients $r_n = \frac{h_n}{h_{n-1}}$. These coefficients are precisely the same as the ones appearing in the recursion relation explicitly constructing orthogonal polynomials (for an even potential as we have in our case)
\be
p_{n+1} (z) = z\, p_n (z) - r_n\, p_{n-1} (z).
\ee
\noindent
Conversely, these recursion coefficients may also be written in terms of the partition function:
\be
\label{randZ}
r_N = \frac{Z (N+1)\, Z (N-1)}{Z (N)^2}.
\ee
\noindent
Of course at the end of the day the partition function \eqref{partitionfunctionV} is solely determined by the potential \eqref{potentialV}, which means that these coefficients must also be determined by this choice of potential. In fact they are; they obey the so-called string equation \cite{biz80}, a finite-difference recursive equation which in the case of the quartic potential is
\be 
\label{eq:stringequation}
r_n \left(1 - \frac{\lambda}{6} \left( r_{n-1} + r_n + r_{n+1} \right) \right) = n g_s.
\ee
\noindent
Note that their relevance extends beyond the above formulae, as these are also the natural objects to address and compute when considering the 't~Hooft large $N$ limit, see \cite{biz80}. Furthermore, these are also the natural quantities to work with when constructing the large $N$ resurgent transseries expansion for the partition function (or for the free energy), \textit{e.g.}, \cite{asv11, sv13}.

\begin{figure}[t!]
\begin{tabular}[b]{cc}
\begin{subfigure}[b]{0.45\columnwidth} 
\includegraphics[width=9.3cm]{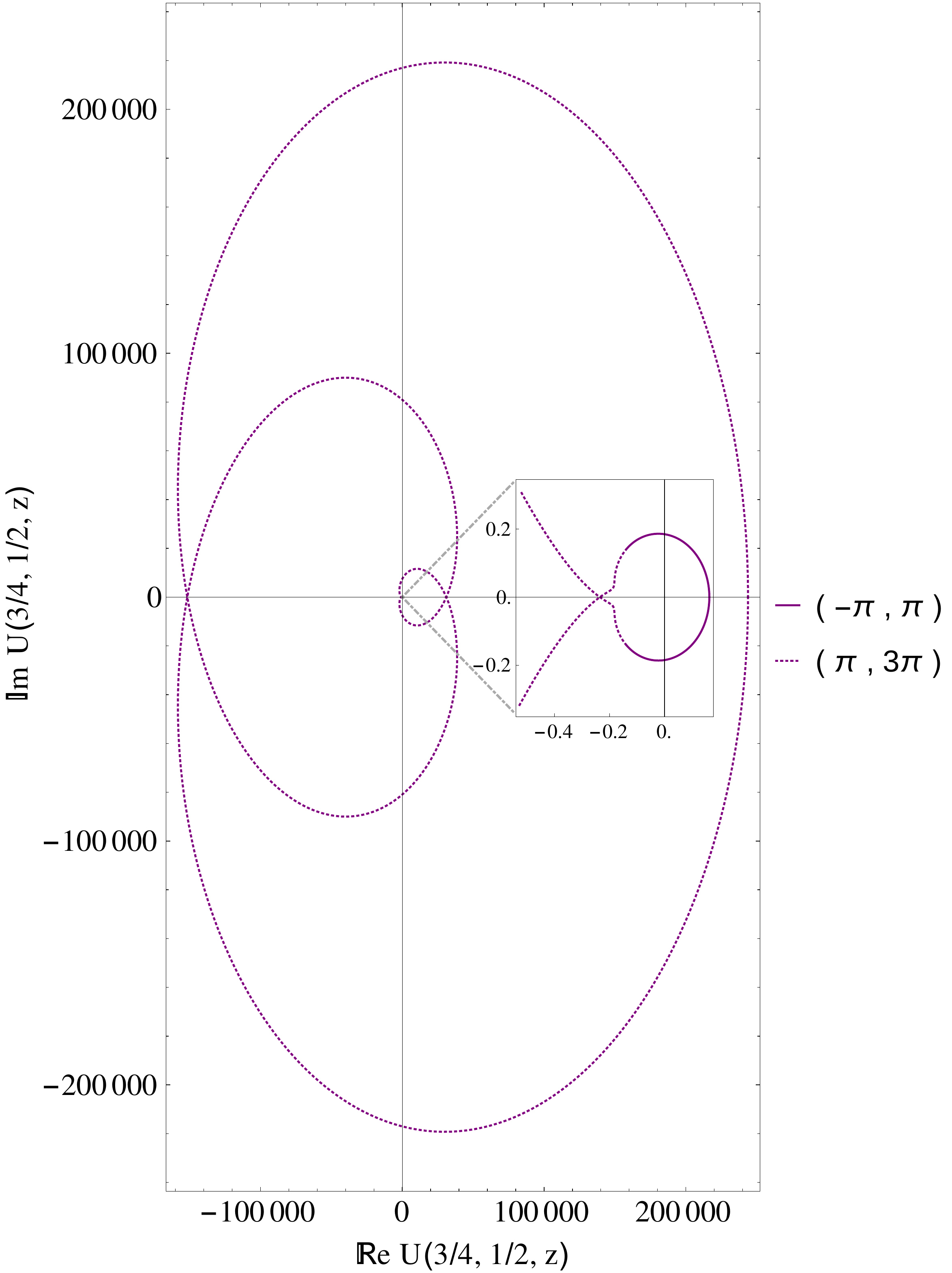}
\end{subfigure}
&
\begin{tabular}[b]{c}
\begin{subfigure}[t]{0.4\columnwidth}
\quad\quad\quad\quad
\includegraphics[width=5.5cm]{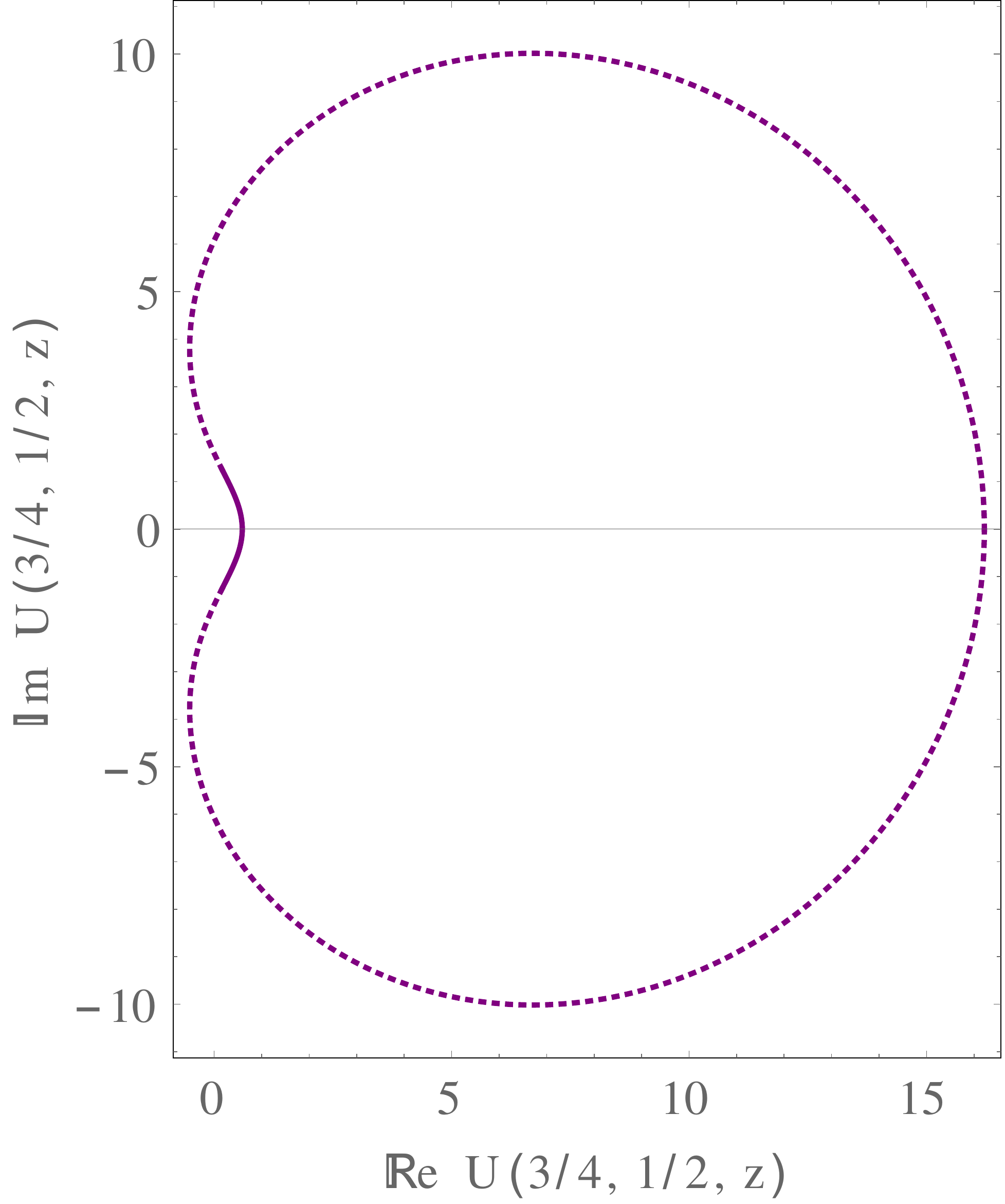}
\end{subfigure}
\\
\begin{subfigure}[b]{0.4\columnwidth}
\quad\quad\quad\quad
\includegraphics[width=5.5cm]{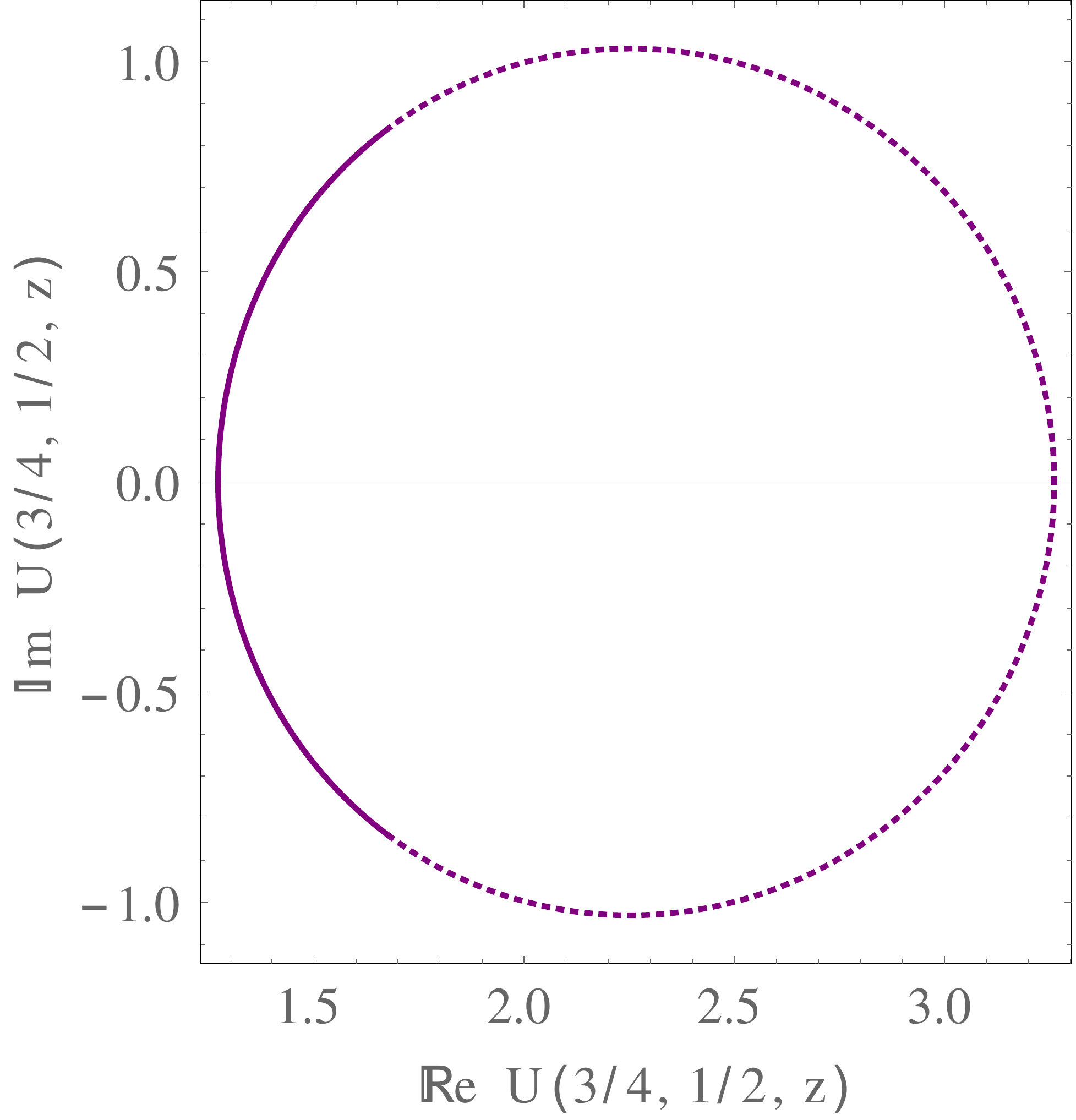}
\end{subfigure}
\end{tabular}
\end{tabular}
\caption{Illustration of the (double) monodromy of the confluent hypergeometric function of the second kind, for fixed values of $a=\frac{3}{4}$ and $b=\frac{1}{2}$, and different values of $|z|=0.1$ (lower right), $|z|=1.1$ (upper right), and $|z|=10.1$ (left). The figures plot $U \left( \left. a,b\, \right| z \right)$ over the complex $z$-plane as $\arg z \in (-\pi,3\pi)$, with the solid line representing the first turn, $\arg z \in (-\pi,\pi)$, and the dotted line the second, $\arg z \in (\pi,3\pi)$. Note the difference in scales between principal and secondary sheets, increasingly significant as $|z|$ grows. In the left plot we have inclosed a zoom-in close to the origin, in order to show the trajectory along the first sheet in more detail.}
\label{fig:confluent_monodromy}
\end{figure}

Using these relatively standard techniques, one may now compute the exact partition function for a few values of the rank $N$. If $N$ is small, one may proceed and compute the polynomial norms, $h_n$, directly. First introduce the moments (under appropriate convergence conditions)
\be
\label{quartic_moments}
\mathfrak{m}_n \equiv \int_{\BR} \rmd \mu (z)\, z^n = \frac{1+(-1)^n}{4\pi}\, \Gamma \left(\frac{n+1}{2}\right) \left( - \frac{6 g_s}{\lambda} \right)^{\frac{n+1}{4}}  \, U \left( \left. \frac{n+1}{4}, \frac{1}{2}\, \right| - \frac{3}{2 \lambda g_s} \right),
\ee
\noindent
in terms of which all results that follow may be expressible. Of course odd moments vanish as the potential we are considering is an even function. The result is expressible in terms of the confluent hypergeometric function of the second kind $U \left( \left. a,b\, \right| z \right)$ (see, \textit{e.g.}, \cite{olbc10}), which has a branch cut along the negative real axis in $z$, $\arg z = \pm\pi$, and integral representation
\be
U \left( \left. a,b\, \right| z \right) = \frac{1}{\Gamma (a)} \int_{0}^{+\infty} \rmd x\, x^{a-1} \left( 1+x \right)^{b-a-1} \rme^{-z\, x},
\ee
\noindent
when $\re\, a > 0$ and $\left| \arg z \right| < \frac{\pi}{2}$. Because of the branch cut, this function has non-trivial monodromy, given by \cite{olbc10}
\be
\label{confluent_monodromy}
U \left( \left. a,b\, \right| \rme^{2\pi\rmi m}\, z \right) = \rme^{-2\pi\rmi b m}\, U \left( \left. a,b\, \right| z \right) + \frac{2\pi\rmi\, \rme^{- \rmi\pi b m}}{\Gamma \left( 1+a-b \right) \Gamma \left( b \right)}\, \frac{\sin \left( \pi b m \right)}{\sin \left( \pi b \right)}\, M \left( \left. a,b\, \right| z \right), \qquad m \in \BZ,
\ee
\noindent
where now $M \left( \left. a,b\, \right| z \right)$ is the confluent hypergeometric function of the first kind; an entire function with integral representation
\be
M \left( \left. a,b\, \right| z \right) = \frac{\Gamma (b)}{\Gamma (a)\, \Gamma (b-a)} \int_{0}^{1} \rmd x\, x^{a-1} \left( 1-x \right)^{b-a-1} \rme^{z\, x},
\ee
\noindent
when $\re\, b > \re\, a > 0$. In our quartic-potential example \eqref{quartic_moments} one always has $b = \frac{1}{2}$, implying from \eqref{confluent_monodromy} that we have to rotate twice around the origin in the complex $z$-plane in order to return to the starting point, as illustrated in figure \ref{fig:confluent_monodromy}. This will also be a distinguishable feature for both recursion coefficients, free energy and partition function of our quartic gauge theory. 

\begin{figure}[t!]
\begin{tabular}[b]{cc}
\begin{subfigure}[b]{0.425\columnwidth}
\includegraphics[width=8.1cm]{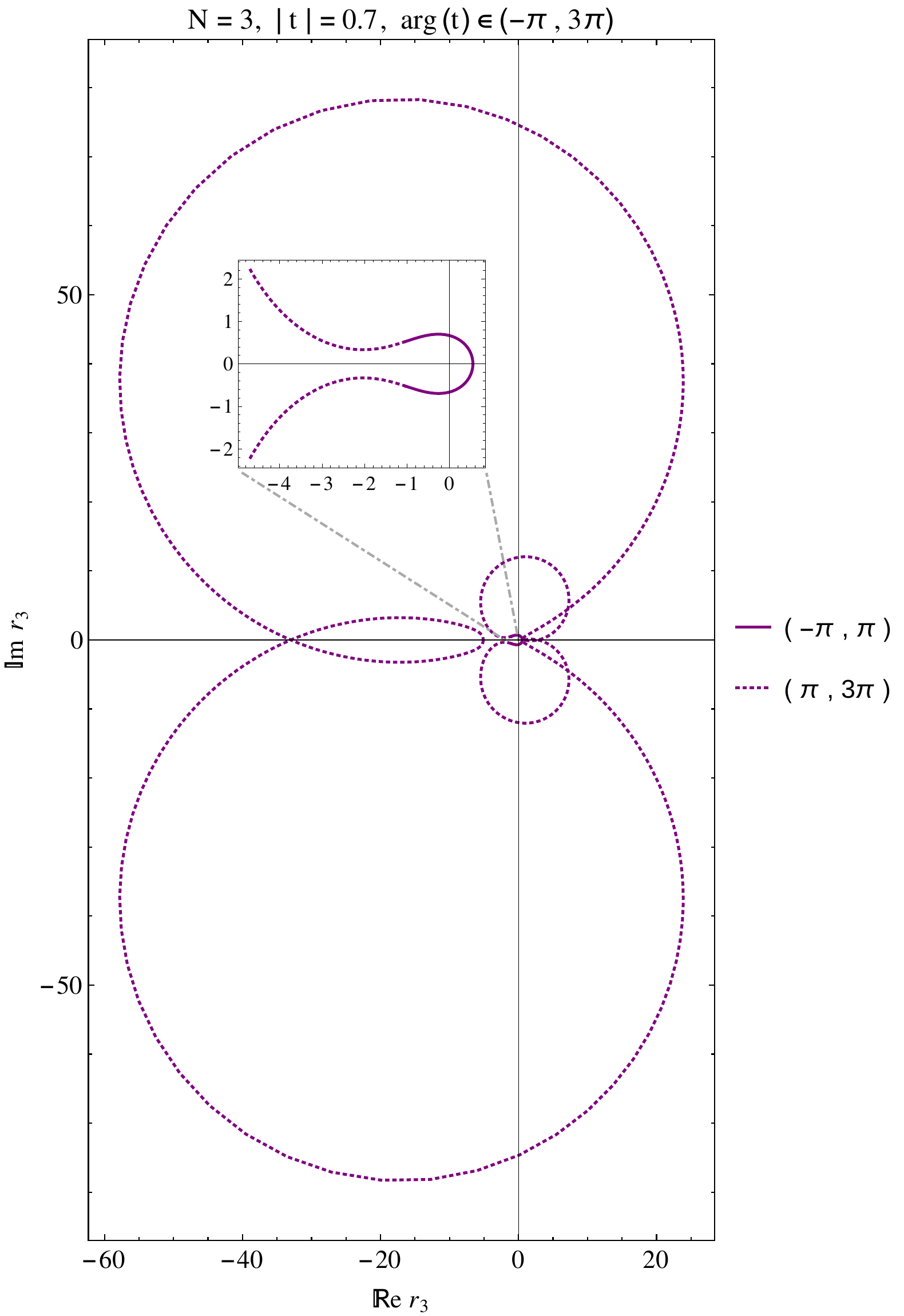}
\end{subfigure}
&
\begin{tabular}[b]{c}
\begin{subfigure}[b]{0.4\columnwidth}
\includegraphics[width=8cm]{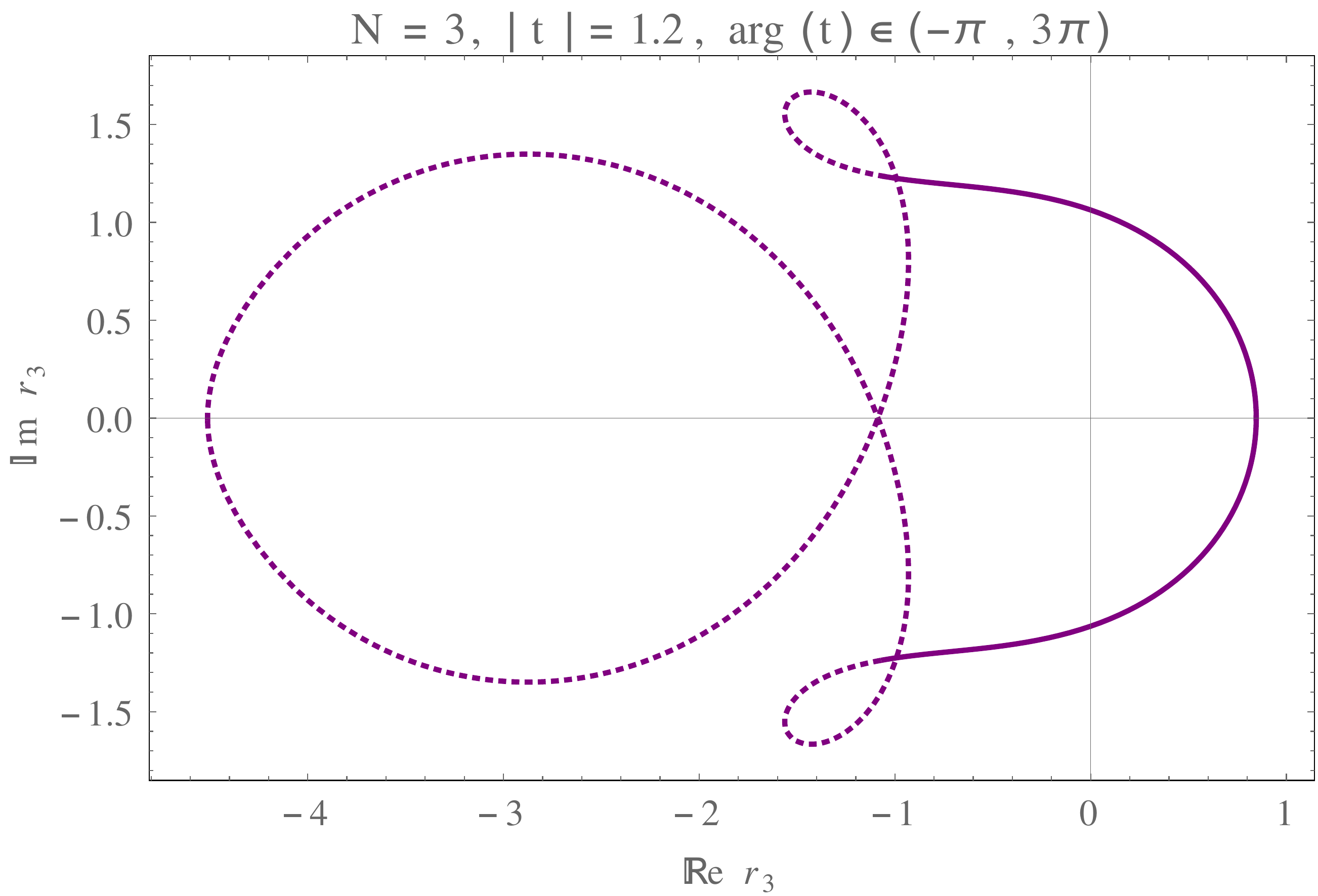}
\end{subfigure}
\\
\\
\begin{subfigure}[b]{0.4\columnwidth}
\includegraphics[width=8cm]{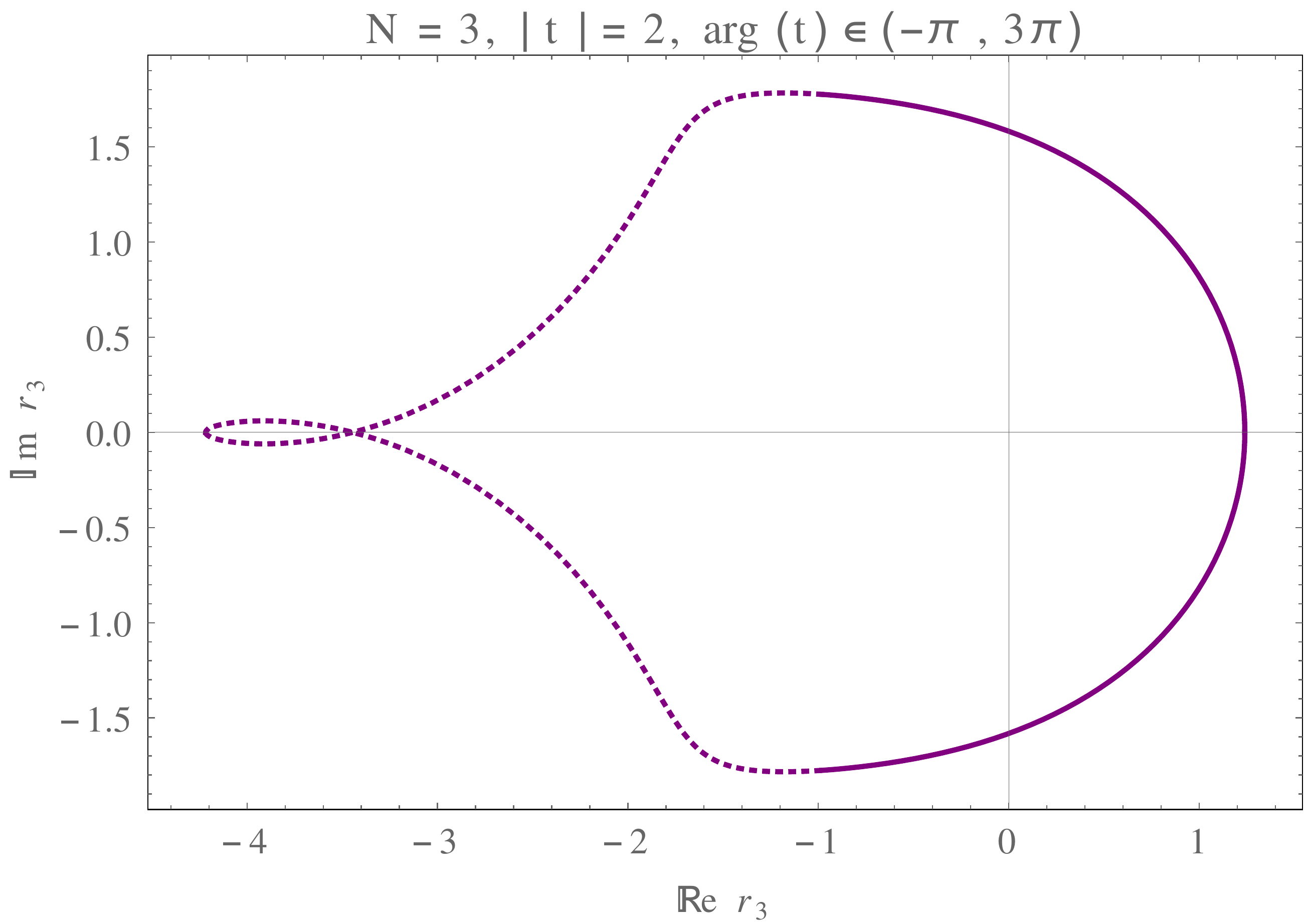}
\end{subfigure}
\end{tabular}
\end{tabular}
\caption{Monodromy of the recursion coefficient $r_3$, for different values of $|t|=0.7$ (left), $|t|=1.2$ (upper right), and $|t|=2$ (lower right). The solid line corresponds to the first sheet, $\arg(t) \in (-\pi, \pi)$, and the dotted line to the second, $\arg(t) \in (\pi, 3\pi)$. In the left plot we inclose a zoom-in close to the origin, to see the trajectory along the first sheet in more detail.}
\label{fig:r3_monodromy}
\end{figure}

\begin{figure}[t!]
\begin{tabular}[b]{cc}
\begin{subfigure}[b]{0.45\columnwidth}
\includegraphics[height=15.3cm]{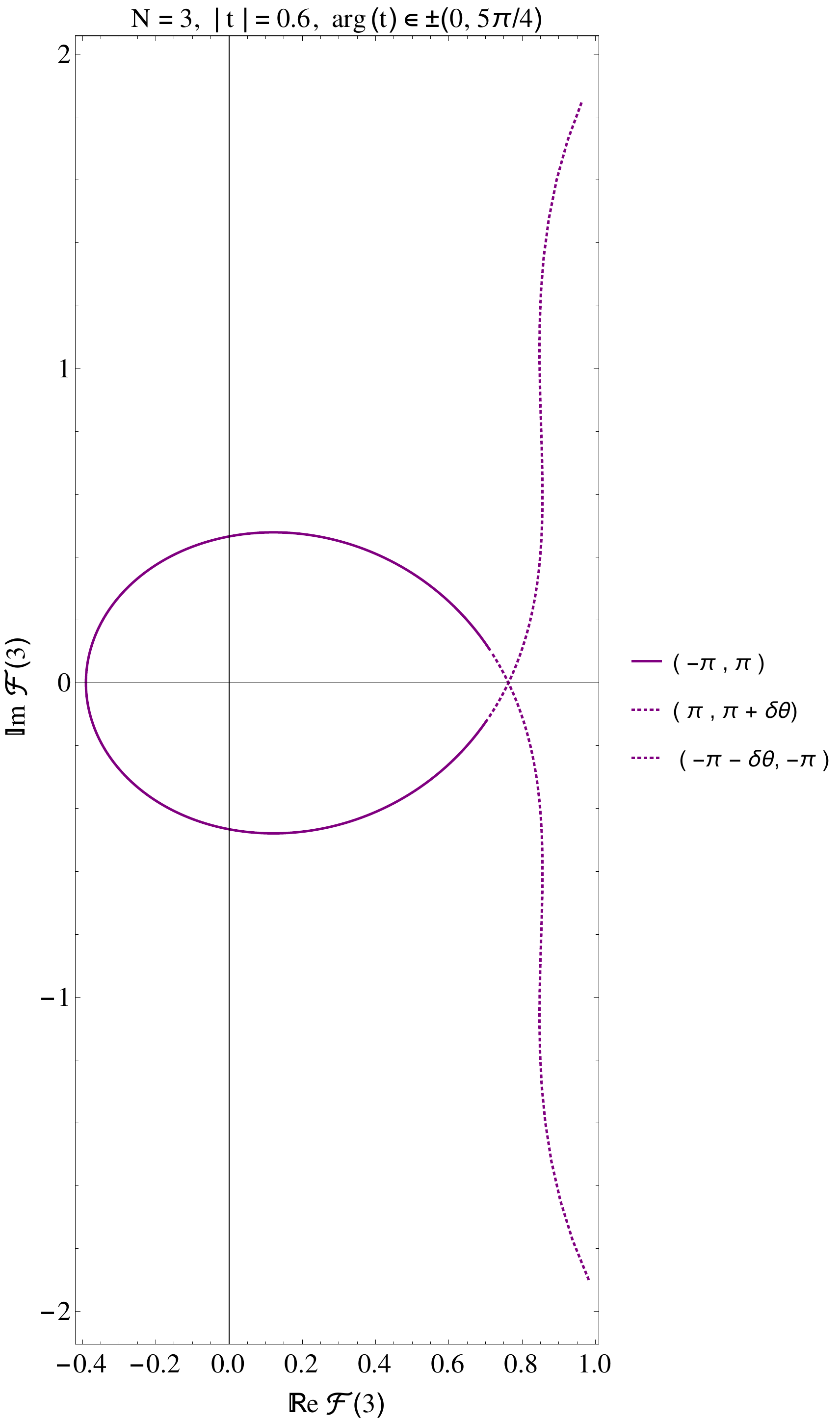}
\end{subfigure}
&
\begin{tabular}[b]{c}
\begin{subfigure}[b]{0.4\columnwidth}
\quad\quad
\includegraphics[height=4.6cm]{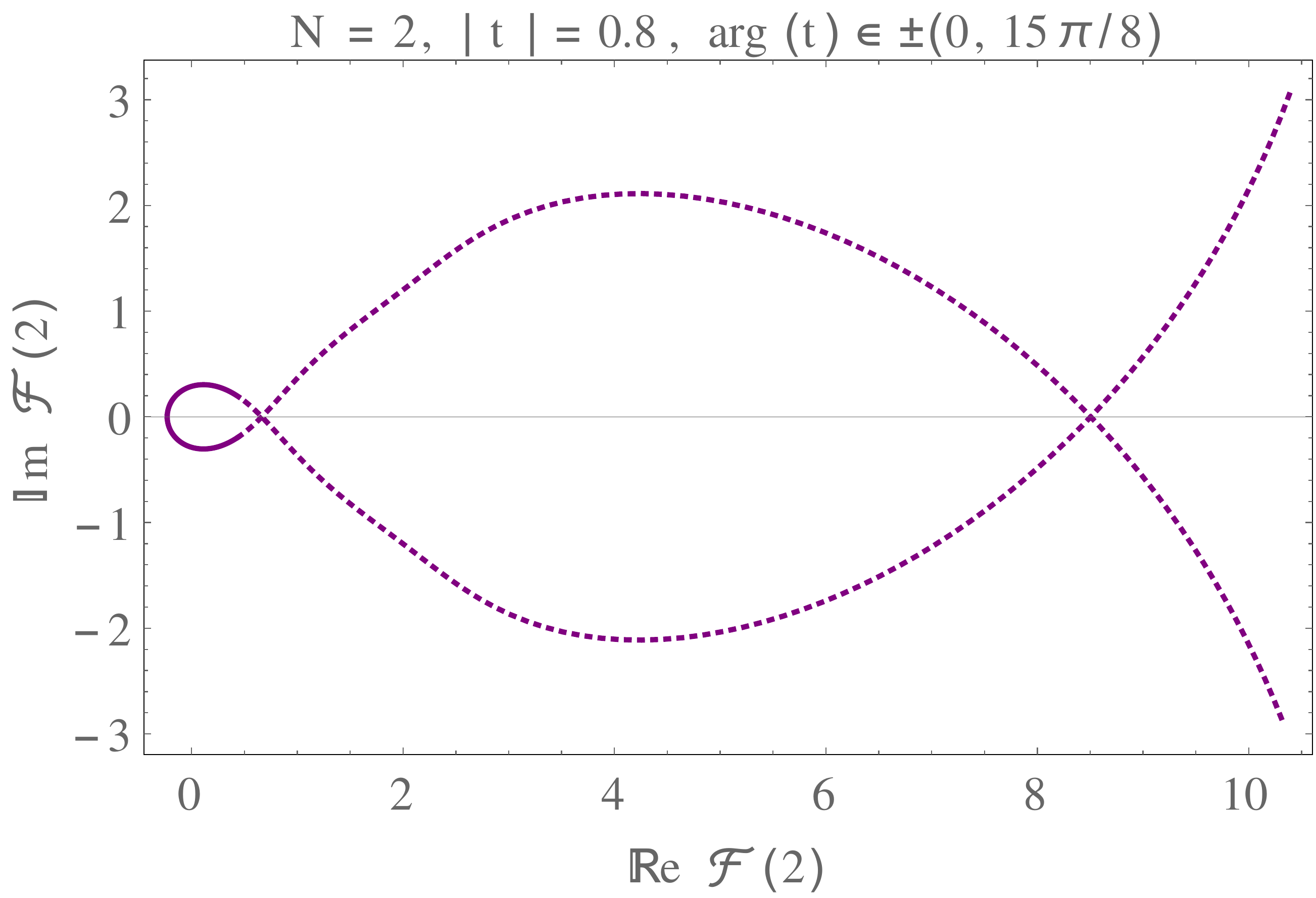}
\end{subfigure}
\\
\\
\begin{subfigure}[b]{0.3\columnwidth}
\quad\quad\quad
\includegraphics[height=10cm]{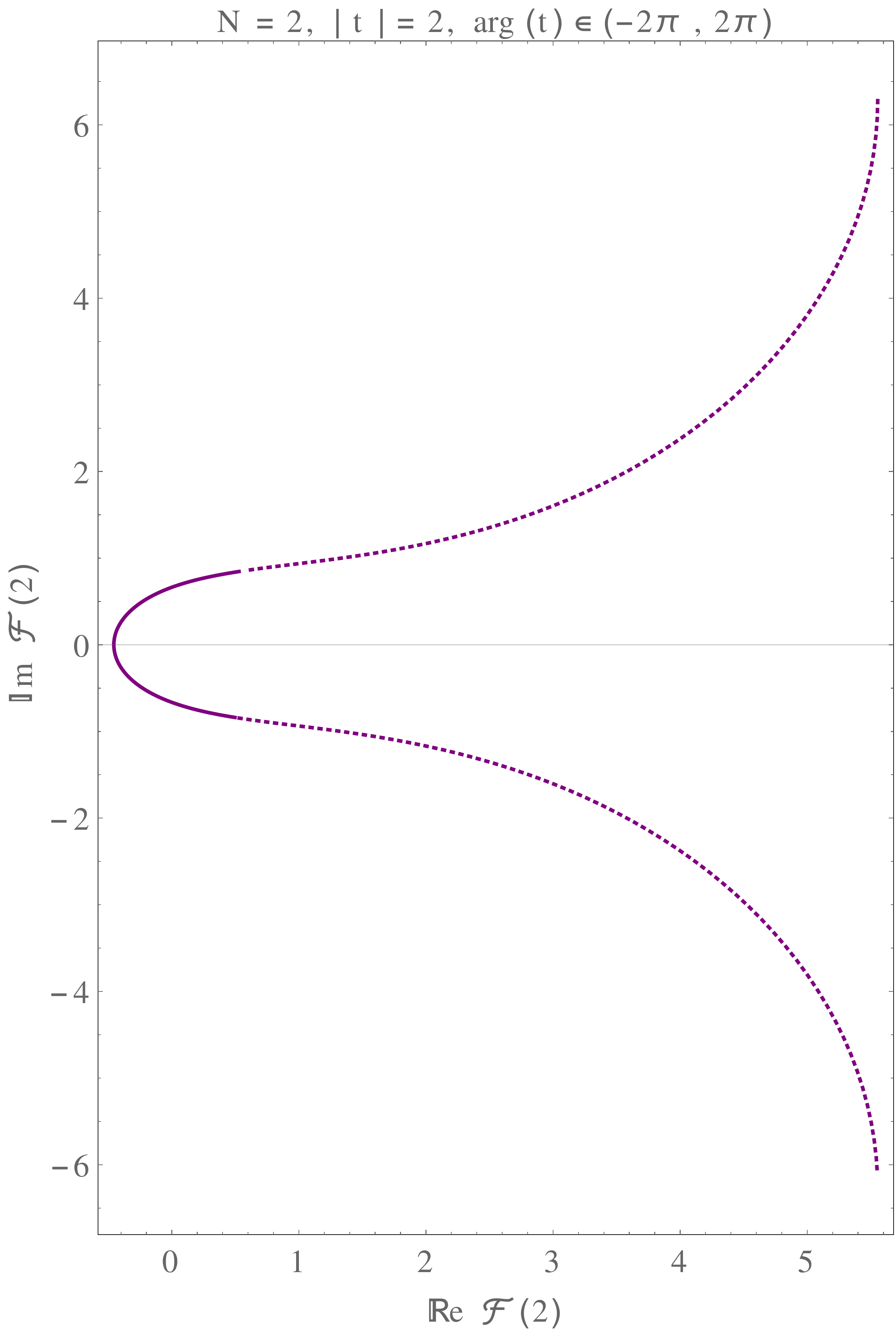}
\end{subfigure}
\end{tabular}
\end{tabular}
\caption{Monodromy of the free energy $\CF(3)$, when $N=3$, for $|t|=0.6$ (left); and of the free energy $\CF(2)$, when $N=2$, for different values $|t|=0.8$ (upper right) and $|t|=0.6$ (lower right). As usual, the solid line corresponds to $\arg(t) \in (-\pi, \pi)$ while the dotted line now corresponds to $\arg(t) \in \pm (\pi, \pi+ \delta\theta)$, where we set $\delta\theta=\pi/4$ (left), $\delta\theta=7\pi/8$ (upper right), and $\delta\theta=\pi$ (lower right). In some plots we have not included the full dotted curves as they show no more relevant features beyond what is displayed (their range gets naturally enlarged by force of the logarithm).}
\label{fig:F23_monodromy}
\end{figure}

One may now explicitly construct the orthogonal polynomials of the quartic matrix model in terms of the moments, $\mathfrak{m}_n$, and find the following first few recursion coefficients
\bea 
\label{r_1}
r_1 &=& \frac{\mathfrak{m}_2}{\mathfrak{m}_0}, \\
\label{r_2}
r_2 &=& \frac{\mathfrak{m}_4}{\mathfrak{m}_2} - \frac{\mathfrak{m}_2}{\mathfrak{m}_0}, \\
\label{r_3}
r_3 &=& \frac{\mathfrak{m}_0}{\mathfrak{m}_2} \frac{\mathfrak{m}_4^2 - \mathfrak{m}_6\, \mathfrak{m}_2}{\mathfrak{m}_2^2 - \mathfrak{m}_4\, \mathfrak{m}_0}, \\
\label{r_4}
r_4 &=& \frac{\mathfrak{m}_8\, \mathfrak{m}_2}{\mathfrak{m}_6\, \mathfrak{m}_2 - \mathfrak{m}_4^2} - \frac{\mathfrak{m}_4\, \mathfrak{m}_2}{\mathfrak{m}_4\, \mathfrak{m}_0 - \mathfrak{m}_2^2} - \frac{\mathfrak{m}_6\, \mathfrak{m}_2\, \big( \mathfrak{m}_6\, \mathfrak{m}_0 - \mathfrak{m}_4\, \mathfrak{m}_2 \big)}{\left( \mathfrak{m}_6\, \mathfrak{m}_2 - \mathfrak{m}_4^2 \right) \left( \mathfrak{m}_4\, \mathfrak{m}_0 - \mathfrak{m}_2^2 \right)}.
\eea
\noindent
Next, it is simple to use these results to find the partition functions at different values of the rank. Notice that while the polynomial norms and the recursion coefficients are, in general, rational functions, we can see from their structure that certain cancellations occur, ensuring that at the end of the day the partition functions are just a sum of products of hypergeometric functions. This could have been predicted from the start, since the Vandermonde determinant is a polynomial. Explicitly, the partition functions for $N=2,3,4,5$ are given by
\bea
\label{Z_2}
Z(2) &=& \mathfrak{m}_2\, \mathfrak{m}_0, \\
\label{Z_3}
Z(3) &=& \mathfrak{m}_2\, \Big( \mathfrak{m}_4\, \mathfrak{m}_0 - \mathfrak{m}_2^2 \Big), \\
Z(4) &=& \Big( \mathfrak{m}_6\, \mathfrak{m}_2 - \mathfrak{m}_4^2 \Big)\, \Big( \mathfrak{m}_4\, \mathfrak{m}_0 - \mathfrak{m}_2^2 \Big), \\
Z(5) &=& \Big( \mathfrak{m}_6\, \mathfrak{m}_2 - \mathfrak{m}_4^2 \Big)\,  \Big\{ \mathfrak{m}_8\, \Big( \mathfrak{m}_4\, \mathfrak{m}_0 - \mathfrak{m}_2^2 \Big) + \mathfrak{m}_6\, \Big( \mathfrak{m}_6\, \mathfrak{m}_0 - \mathfrak{m}_4\, \mathfrak{m}_2 \Big) - \nonumber \\
&&
- \mathfrak{m}_4\, \Big( \mathfrak{m}_6\, \mathfrak{m}_2 - \mathfrak{m}_4^2 \Big) \Big\}.
\label{Z_5}
\eea

Note that for different values of $N$ the above partition functions are not simply numbers, but \textit{functions}. In these formulae they are implicitly functions of the couplings in the matrix integral via \eqref{quartic_moments}, which we trade for the 't~Hooft coupling $t$ as usual. As a consequence of \eqref{confluent_monodromy}, and as we shall now illustrate, these functions display intricate monodromy structures. For instance, in figure \ref{fig:r3_monodromy} we illustrate the monodromy properties of $r_3$, for three different values of $|t|$ and varying $\arg (t)$. Furthermore, the free energy\footnote{A word on notation and normalizations: we shall follow standard practice of normalizing the partition function with respect to the Gaussian weight \eqref{gaussian_Z}, which means in practice the partition function is in fact given by
\be
\CZ = \frac{Z}{Z_{\text{G}}}.
\ee
\noindent
However, with a slight abuse of notation and following common practice in earlier papers, this will always be implicitly assumed so that whenever we refer to $Z$ this is always the normalized result $\CZ$. The free energy will also always be normalized as $\CF = \log \CZ$, but in this case we shall keep it explicit in notation.} also displays distinctive monodromy features. The one novelty is that due to the logarithm the scale in the second sheet may be considerably larger than the corresponding one in the first sheet. This is illustrated in figure \ref{fig:F23_monodromy} for different values of $N$ and $t$ (in fact in these plots we do not show the entire range corresponding to the second sheet, in order to keep the pictures small enough for illustration purposes). Note that due to the logarithm, now the curves do not close upon themselves. 

It is clear that the monodromy features displayed in figures \ref{fig:r3_monodromy} and \ref{fig:F23_monodromy} are quite nontrivial, and they change as we change the value of $N$. What is important to have in mind is that all these features must be precisely captured by the large $N$ transseries which will be introduced in the following section. Furthermore, there is one single transseries describing the quartic matrix model, and this \textit{one} transseries must be able to reproduce \textit{all} these monodromies for \textit{all} different values of $N$. As we shall see, this will be remarkably captured by instanton physics.

\section{Finite $N$ from Resurgent Large $N$}\label{sec:resurgent}

The construction of the perturbative $1/N$ expansion has a long history. In the matrix model context it starts with a (planar) spectral curve \cite{bipz78}, out of which one may recursively construct the full perturbative series---an endeavor which started out in \cite{ackm93} and culminated in \cite{eo07}. But, as we shall now discuss, there is much more to gauge theory at finite $N$ beyond its perturbative $1/N$ expansion. Depending on the values of the parameters, and on Stokes phenomenon, instantons may be crucial to achieve exponential accuracy in some results, or, instead, they become exponentially enhanced rather than suppressed to completely change the perturbative results and \textit{correctly} reproduce many of the intricate monodromy features we have discussed in the previous section. So, how are instantons incorporated into the large $N$ expansion?

As already explained in the introduction, this is achieved via transseries which go beyond the large $N$ expansion by including all its nonperturbative corrections. In practice one deals with a formal expression, such as \eqref{initial_transseries}. But it is important to notice that such transseries contain as much information as a would-be analytic expression for whatever function we are trying to describe. The explicit connection between the two is achieved by the resummation of the former into the latter. This is actually similar to the role that Taylor power-series play in describing or representing entire functions. In our case, while similar in spirit, the ``power-series game'' becomes a little bit harder in practice due to the existence of many singularities and branch cuts. The prominent role of transseries thus comes about, since finding analytic solutions is doomed to fail for most problems, while transseries representations yield natural completions to na\"{i}ve perturbative approaches. These in fact include everything needed for a complete nonperturbative description of the solutions. Furthermore, treating parameters such as the rank $N$ as formal variables removes any previous constraints on their domain of validity (\textit{e.g.}, $N$ must be an integer), so that after resummation they can even take real or complex values. 

The observables we shall focus upon are the partition function and the free energy. Let us start with the free energy, $\CF (N,t)$. In our example of the quartic matrix model, the construction of a large $N$ transseries representation for the free energy starts by addressing the coefficients $r_N$ (and we shall also display results for these). The large $N$ transseries for $r_N$ is a resurgent function of the form
\be
\label{Rtransseries}
\CR(N,t) = \sum_{n=0}^{+\infty} \sigma^n\, \rme^{- n N \frac{A(t)}{t}}\, \sum_{g=0}^{+\infty} N^{-g-\beta_n}\, t^{g+\beta_n} R^{(n)}_g (t),
\ee
\noindent
which solves the continuous version of the string equation \eqref{eq:stringequation} when expressed in terms of the 't~Hooft coupling. We refer to appendix \ref{sec:appendix}, and references therein, for a more detailed explanation of this transseries solution, \textit{e.g.}, the choice of variables, the explicit instanton action, the coefficients $\beta_n$, the maximum orders up to which we have computed the $g$-coefficients, and a few explicit examples. Let us nonetheless stress a few points: there are two distinct sums, one in the instanton number, $n$, and the other in the perturbative order, $g$; the rank $N$ is treated as a (continuous) formal parameter; at large $N$ the nonperturbative exponential contributions are controlled by the ratio $A(t)/t$ involving the matrix model instanton action \eqref{eq:instantonaction}; the transseries parameter $\sigma$ is so far arbitrary and needs to be fixed for any numerical evaluations; the perturbative coefficients $R^{(n)}_g(t)$ grow factorially fast with $g$, turning each instanton series in $1/N$ asymptotic; and that all building blocks of the transseries are in fact \textit{functions} of the 't~Hooft coupling $t$. Having determined the transseries for the coefficients $r_N$, one then uses equation \eqref{TodaRF} in order to obtain the transseries for the free energy,
\be
\label{Ftransseries}
\CF(N,t) = \sum_{n=0}^{+\infty} \sigma^n\, \rme^{- n N \frac{A(t)}{t}}\, \sum_{g=0}^{+\infty} N^{-g-\beta^\CF_n}\, t^{g+\beta^\CF_n} \CF^{(n)}_g (t).
\ee
\noindent
We once more refer to the appendix for more technical details on this transseries construction. Note that when addressing any other gauge theory, the starting point and the method that now follows will be exactly the same. The only difference would be a distinct instanton action and multi-instanton perturbative coefficients, possibly computed diagramatically, \textit{i.e.}, the difference would essentially amount to having a distinct content in the appendix of the paper.

One thing to notice is that expressions \eqref{Rtransseries} or \eqref{Ftransseries} are not the most general transseries solutions to the quartic matrix model; see \cite{asv11}. Instead, it turns out that the instanton action has a symmetric companion of opposite sign which also solves the relevant differential equation. Consequently, the complete transseries solution will depend upon two parameters and it will include logarithmic monomials in $N$ associated to resonant sectors. While knowledge of this complete transseries is necessary in order to understand the resurgence properties of the free energy---namely, how the coefficients of one sector grow and relate to coefficients from other sectors---it will not be needed for the particular resummations we address in the present paper. 

Finally, one still has to fix $\sigma$ in order to obtain any numbers at the end of the day. Typically, a transseries with a given fixed choice of $\sigma$ will be valid in a specific wedge of the coupling-constant complex-plane; in this case of the complex plane associated to the 't~Hooft parameter $t$. Different wedges are separated by singular directions on the Borel plane, known as \textit{Stokes lines}. This is where the multi-instanton singularities lie. For our present problem, these singular directions on the Borel plane are located at either $\theta=0$ or $\theta=\pi$ \cite{asv11, as13}, but for the problems we shall address in the following we only need to consider the $\theta=0$ case. Crossing this Stokes line implies that the transseries parameter will ``jump'' or ``turn on'', in the sense that any exponentially suppressed contributions previously neglected (as they were invisible behind the perturbative expansion) must now be taken into account as they will start growing and eventually may take dominance. To make this concrete, consider the free energy transseries \eqref{Ftransseries} and write it explicitly as $\CF(N,t,\sigma)$. Stokes phenomena then translates to the statement\footnote{In the language of resurgence this jump is captured by the so-called Stokes automorphism; see, \textit{e.g.}, \cite{asv11, as13}.}
\be
\label{Stokes}
\CS_{0^+} \CF(N,t,\sigma) = \CS_{0^-} \CF(N,t,\sigma+S_1),
\ee
\noindent
where the resummation $\CS_{\theta}$ was defined in \eqref{resummation_definition}, and where the shift in the transseries parameter is controlled by the Stokes constant $S_1$. A particular case of \eqref{Stokes} is when we start off with $\sigma=0$, meaning the transseries has just the perturbative component, and then after crossing the Stokes line the nonperturbative contributions are turned on with a parameter $\sigma=S_1$.

Now, as we want to bring together and compare the two representations of, say, the free energy, the one obtained analytically and the one obtained via resurgent transseries---in fact to show that they are equal at integer $N$---we need to resum the formal transseries into a function. Because the transseries is a double sum we must undergo a two-step process which bears the name of Borel--\'Ecalle resummation. The first name deals with each of the asymptotic series in $1/N$, while the second takes care of the sum over multi-instantons. Both are important, in solving different problems, but, in practice, when we need to get a number out of a particular example the Borel resummation takes most of the attention. Let us fix an instanton sector, $n$, and consider the asymptotic series
\be
\CF^{(n)} (N,t) \simeq \sum_{g=0}^{+\infty} N^{-g-\beta^\CF_n}\, t^{g+\beta^\CF_n} \CF^{(n)}_g (t).
\ee
\noindent
As mentioned in the introduction for the perturbative case, Borel resummation first computes the Borel transform of the asymptotic series $\CF^{(n)}(N,t)$---a convergent series in $s$ which may be analytically continued---and then evaluates its Laplace transform, yielding the resummation
\be
\label{eq:Bresum}
\CS_\theta \CF^{(n)} (N,t) = \int_0^{\rme^{i\theta}\infty} \rmd s\, \CB [\CF^{(n)}] (s,t)\,\rme^{-s N}.
\ee
\noindent
The choice of the angle $\theta$ for the integration contour must be made carefully due to the singularities of the Borel transform. Once this is done and all perturbative and multi-instanton asymptotic series have been dealt with, one may take the second step and address the sum over multi-instantons. This is immediate, so the Borel--\'Ecalle resummation of the transseries is, finally,
\be
\label{eq:BEresum}
\CS \CF(N,t) = \sum_{n=0}^{+\infty} \sigma^n\, \rme^{- n N \frac{A(t)}{t}}\, \CS_\theta \CF^{(n)}(N,t).
\ee
\noindent
Note that the left-hand-side is nontrivially independent of $\theta$; all one now has to take into account are the Stokes jumps \eqref{Stokes}, \textit{i.e.}, keep track of which wedge in the complex plane are we on.

When we turn to implement equations \eqref{eq:Bresum} and \eqref{eq:BEresum} in an explicit example, such as ours for the quartic free energy, it is often the case that we  cannot perform the Borel transform analytically. This is simply because in most problems nonlinearity prevents us from obtaining closed-form expression for the asymptotic coefficients and only a finite number of such coefficients are available for computation. The standard approach to circumventing this problem is found in using Pad\'e approximants to mimic the analytic continuation of the Borel transform (see, \textit{e.g.}, \cite{bo78, olbc10}). Because the Pad\'e approximant is a rational function of $s$ we can capture some of the Borel singular behavior. Thus, the Borel--Pad\'e resummation provides a numerical implementation of the exact Borel resummation. As such we will define the \textit{Borel--Pad\'e--\'Ecalle} resummation of a transseries, up to the $i$-th instanton sector, as\footnote{To summarize, we are using $S$ and $\textrm{BP}_{\ell}$ to define the numerical approximations to $\CS$ and $\CB$, respectively.}
\bea
\label{eq:BPEresum}
S_\theta^{(i)} \CF(N,t) &=& \sum_{n=0}^{i} \sigma^n\, \rme^{- n N \frac{A(t)}{t}}\, S_\theta \CF^{(n)}(N,t), \\
\label{eq:BPresum}
S_\theta \CF^{(n)}(N,t) &=& \int_0^{\rme^{i\theta} s_{\max}} \rmd s\, \textrm{BP}_{\ell}  [\CF^{(n)}] (s,t)\, \rme^{-s N}.
\eea
\noindent
In the last definition $\textrm{BP}_{\ell} [\CF]$ denotes a (diagonal) order-$\ell$ Pad\'e approximant of the Borel transform, and the numerical integration has a cut-off at $s_{\max}$. The ``\'Ecalle step'' of the resummation is also truncated in practice, since only a few instanton terms are computed. As we shall see, this will nonetheless be more than enough to show the capabilities of transseries resummation. 

\begin{figure}[t!]
\begin{center}
\includegraphics[width=0.48\textwidth]{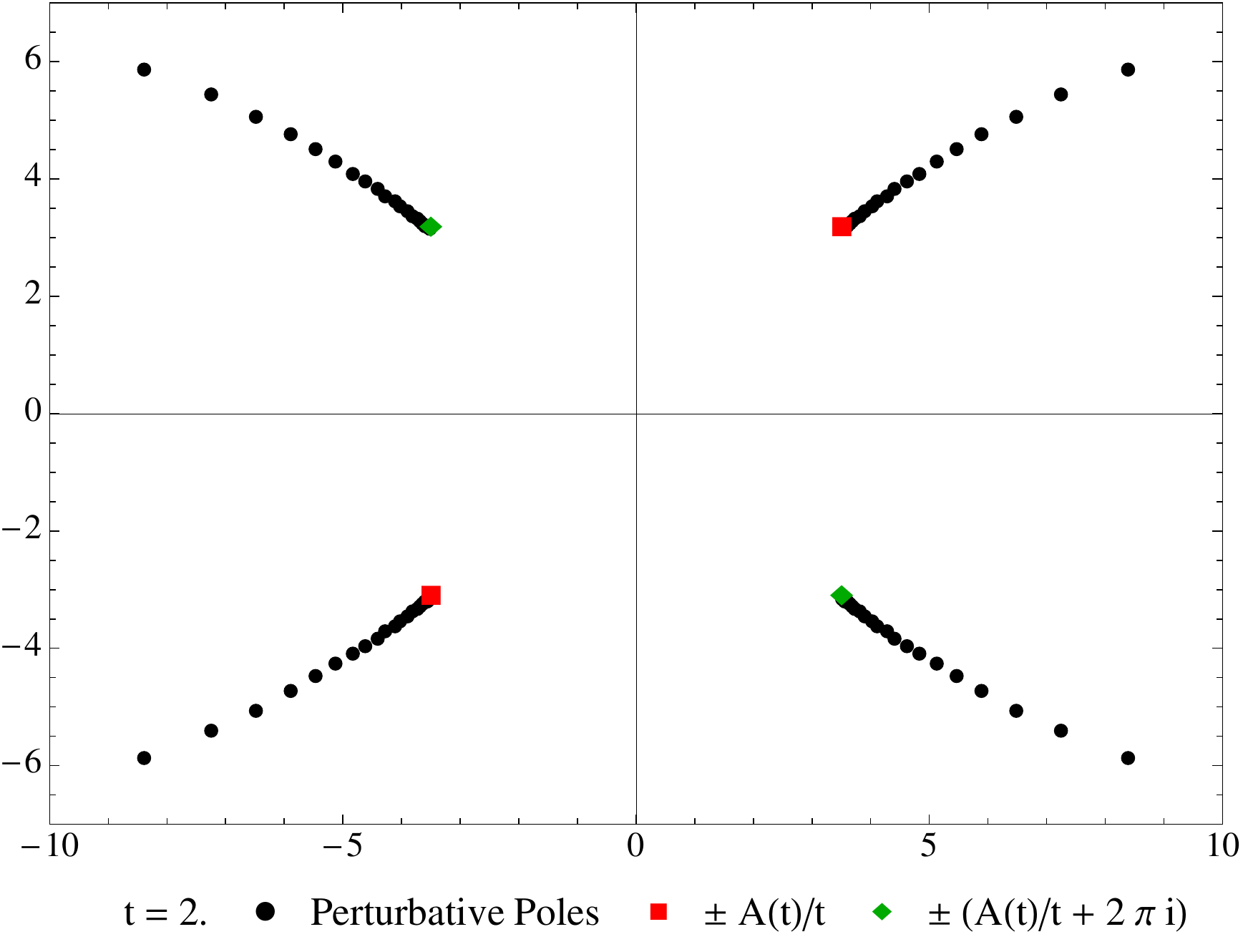}
\hspace{0.02\textwidth}
\includegraphics[width=0.48\textwidth]{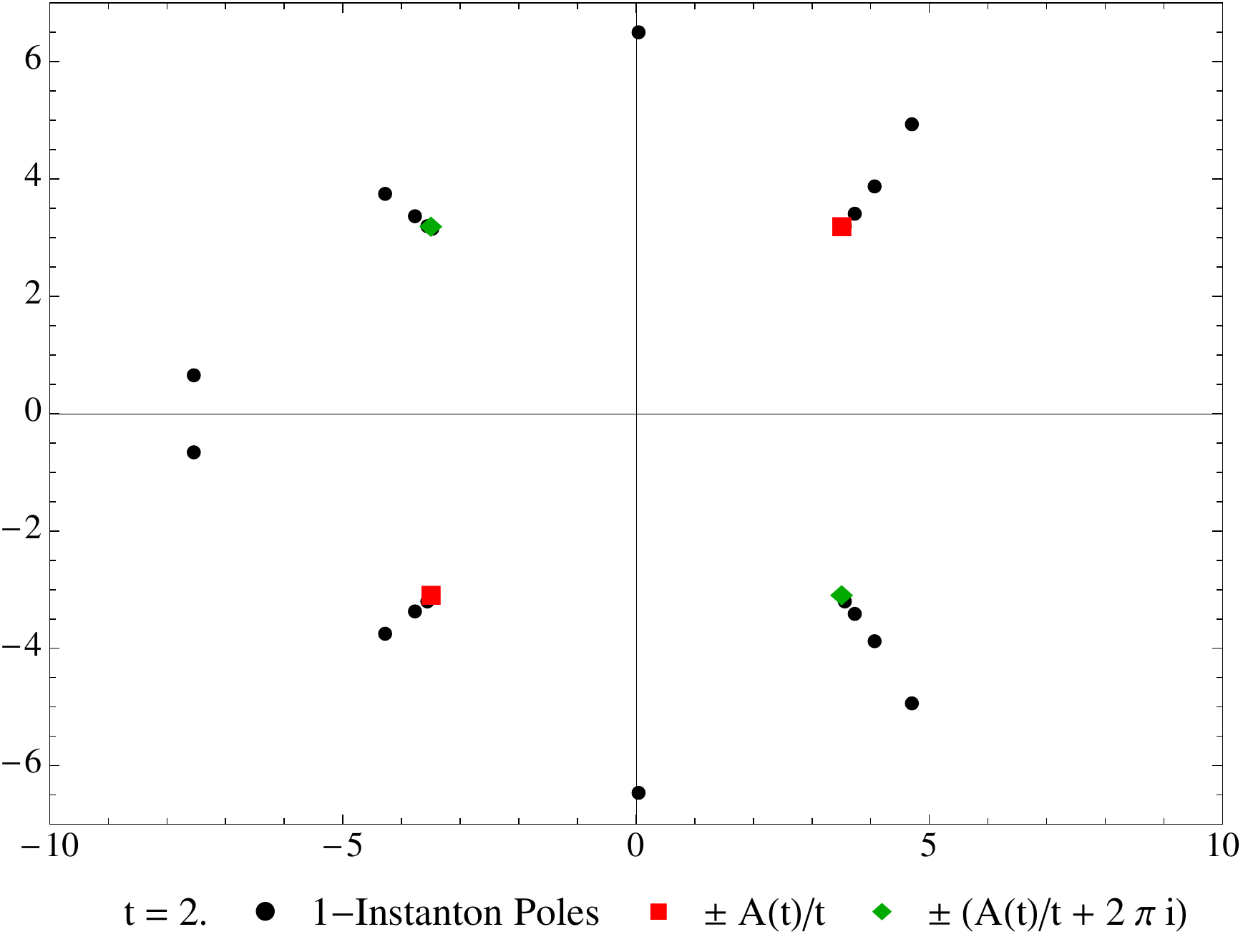} 
\end{center}
\vspace{-1\baselineskip}
\caption{Approximate complex Borel $s$-planes for the perturbative (left, $\CF^{(0)}$) and one-instanton (right, $\CF^{(1)}$) free energies obtained by plotting poles of the Pad\'e approximant when $t=2$. Due to limited computational resources we have less points for higher instanton sectors, as compared to the perturbative sector. Still, the instanton action singularities are very clear, with the accumulation of poles signaling their associated logarithmic branch cuts.}
\label{fig:borelplanesF}
\end{figure}

Note that what we have described is \textit{not} a numerical \textit{method}, but rather a numerical \textit{approximation} to an analytical procedure. As we already mentioned, the Borel--\'Ecalle representation of a function is somewhat analogous to the Taylor power-series representation of (another) function, and this is what \eqref{eq:Bresum} and \eqref{eq:BEresum} set up, out of the transseries. Of course if one is to extract a \textit{number} out of any of these analytical representations, some approximation (or truncation) must be considered. In Taylor power-series one just truncates at a given order and then sums. In the present Borel--\'Ecalle framework, where the functional complexity is larger, one needs to implement the above Borel--Pad\'e--\'Ecalle resummation.

We can see an example in figure \ref{fig:borelplanesF}. The singularities on the Borel--Pad\'e plane convey the image of branch cuts, where the branch points are given by the instanton actions that appear in the transseries. As we have commented before, instanton actions come in pairs of opposite signs. In here, we also notice the presence of a displacement of the instanton action $A(t)/t$ by a constant term $2\pi\rmi$. This is in agreement with general expectations of \cite{dmp11}, where instanton actions in matrix models and topological strings should be linear combinations of spectral curve B-periods, such as in \cite{msw07, msw08}, with spectral curve A-periods, such as in \cite{ps09} (\textit{i.e.}, the factor of $2\pi\rmi t$). Note however that due to the nature of the string equation, the sector associated with this other action is indistinguishable from that of $A(t)/t$ alone, and the two can be combined. In fact if we were to consider a two-parameter transseries\footnote{This should not be confused with the two-parameter transseries in \cite{asv11}, where the two actions are $\pm A$. The inclusion of this second action $\widetilde{A} = A \pm 2\pi\rmi t$ in that set-up would lead to a four-parameter transseries.} with sectors associated to $A$ and $\widetilde{A} = A \pm 2\pi\rmi t$, denoted by $[n|\widetilde{n}]$, then after plugging such an \textit{ansatz} into the string equation we would find that the ``mixed'' coefficients satisfy
\be 
R^{[n|\widetilde{n}]}_g = \frac{(n+\widetilde{n})!}{n!\, \widetilde{n}!} R^{[n+\widetilde{n}|0]}_g.
\ee
\noindent
Plugging this result back into its two-parameter transseries would reduce it to the one-parameter transseries \eqref{Rtransseries}, with a simple shift in parameter $\sigma \to ( 1 + \rme^{\pm 2 \pi \rmi N} ) \sigma$. In the case of integer $N$ we have considered so far, the factor is equal to $2$. But since the transseries parameter $\sigma$ needs to be fixed in any case (which we will do next), the effect of the second instanton action with the $2\pi\rmi$ displacement is already automatically included in our results.

The final issue to address concerns the fact that equations \eqref{Rtransseries} and \eqref{Ftransseries} are actually representing a \textit{family} of transseries, indexed by the transseries parameter $\sigma$. The resummation procedure cannot be complete until $\sigma$ is specified. As we have discussed earlier, its particular value is subject to Stokes transitions that may add or subtract the Stokes constant $S_1$, but it still needs to be fixed at some point, with Stokes transitions then specifying it wherever else. In the present section we are focusing upon the case $t \in \BR^+$, where one can numerically check that the transseries parameter is
\be
\label{valuesigma}
\sigma = \rmi\, \sqrt{\frac{3}{\pi (-\lambda)}}, 
\ee
\noindent
equal to the Stokes constant $S_1$ \cite{msw07, m08, asv11}. Let us recall that $\lambda$ is the quartic coupling-constant which we are going to set to $-1$ without loss of generality. More complicated gauge theories in more ``physical'' scenarios may eventually require that the fixing of transseries parameters must be done against some laboratory measurement.

Now that all the ingredients are on the table, we can explicitly show how adding more and more instanton contributions of the transseries gets us closer and closer to the exact result. In table \ref{tab:transseriesresum} we display explicit numerical examples for the recursion coefficients (left) and the free energy (right). The first four rows correspond to the Borel--Pad\'e resummations \eqref{eq:BPresum} of each sector, the fifth row is their sum \eqref{eq:BPEresum}, and the sixth has the analytical results, \eqref{r_3} and the logarithm of \eqref{Z_3}. In the total result we have labelled with different colors\footnote{This is in fact the color code we shall be using in all subsequent plots.} the digits of the exact result that are matched after including the perturbative (blue), $1$-instanton (green), $2$-instanton (yellow) and $3$-instanton (red) sectors. One can clearly see how the instanton contributions\footnote{There is a slight abuse of notation in the column titles when we write them as $S_0 \CR^{(i)}$ and $S_0 \CF^{(i)}$. What we show is actually the whole contribution to the resummed objects, as in \eqref{eq:BPEresum}, so that herein $S_0 \CR^{(i)}$ and $S_0 \CF^{(i)}$ are in fact multiplied by the relevant powers of $\sigma$ and $\exp(-N A(t)/t)$.} provide exponentially small corrections with respect to the perturbative contribution, where the size of the correction is naturally controlled by the instanton action.

\begin{table}[h!]
\begin{center}
\begin{tabular}{rccc}
Sector & \hspace{0.0cm} & $S_0 \CR^{(n)}$ \quad\quad & $S_0 \CF^{(n)}$ \\[3pt]
\hline 
& & &\\[-8pt]
Perturbative && $2.615\,796\,570\,569\,705\,50\ldots$ \quad\quad & $-1.973\,899\,279\,493\,161\,74\ldots$ \\
1-Instanton  && $0.000\,487\,953\,495\,567\,22\ldots$ \quad\quad & $-0.000\,020\,359\,080\,917\,15\ldots$ \\
2-Instanton  && $0.000\,000\,009\,807\,788\,15\ldots$ \quad\quad & $-0.000\,000\,000\,300\,789\,88\ldots$ \\
3-Instanton  && $0.000\,000\,000\,000\,245\,38\ldots$ \quad\quad & $-0.000\,000\,000\,000\,004\,71\ldots$ \\
& & &\\[-10pt]
Total && $\textcolor{pert}{2.61}\textcolor{1inst}{6\,284\,53}\textcolor{2inst}{3\,873}\,\textcolor{3inst}{306\,2}7\ldots$ \quad\quad & $\textcolor{pert}{-1.973}\,\textcolor{1inst}{919\,638}\,\textcolor{2inst}{874\,87}\textcolor{3inst}{3\,50}\ldots$ \\
& & &\\[-10pt]
Exact && $2.616\,284\,533\,873\,306\,26\ldots$ \quad\quad & $-1.973\,919\,638\,874\,873\,50\ldots$
\end{tabular}
\end{center}
\caption{Comparison of the truncated $\CR$ and $\CF$ transseries, up to the instanton sector $n = 0,1,2,3$, against the exact result for $t = 6$ and $N = 3$. All digits displayed are stable. On the left table the last digit of the resummation must be corrected by the 4-instanton contribution.}
\label{tab:transseriesresum}
\end{table}

\begin{figure}[t!]
\begin{center}
\includegraphics[width=7.75cm]{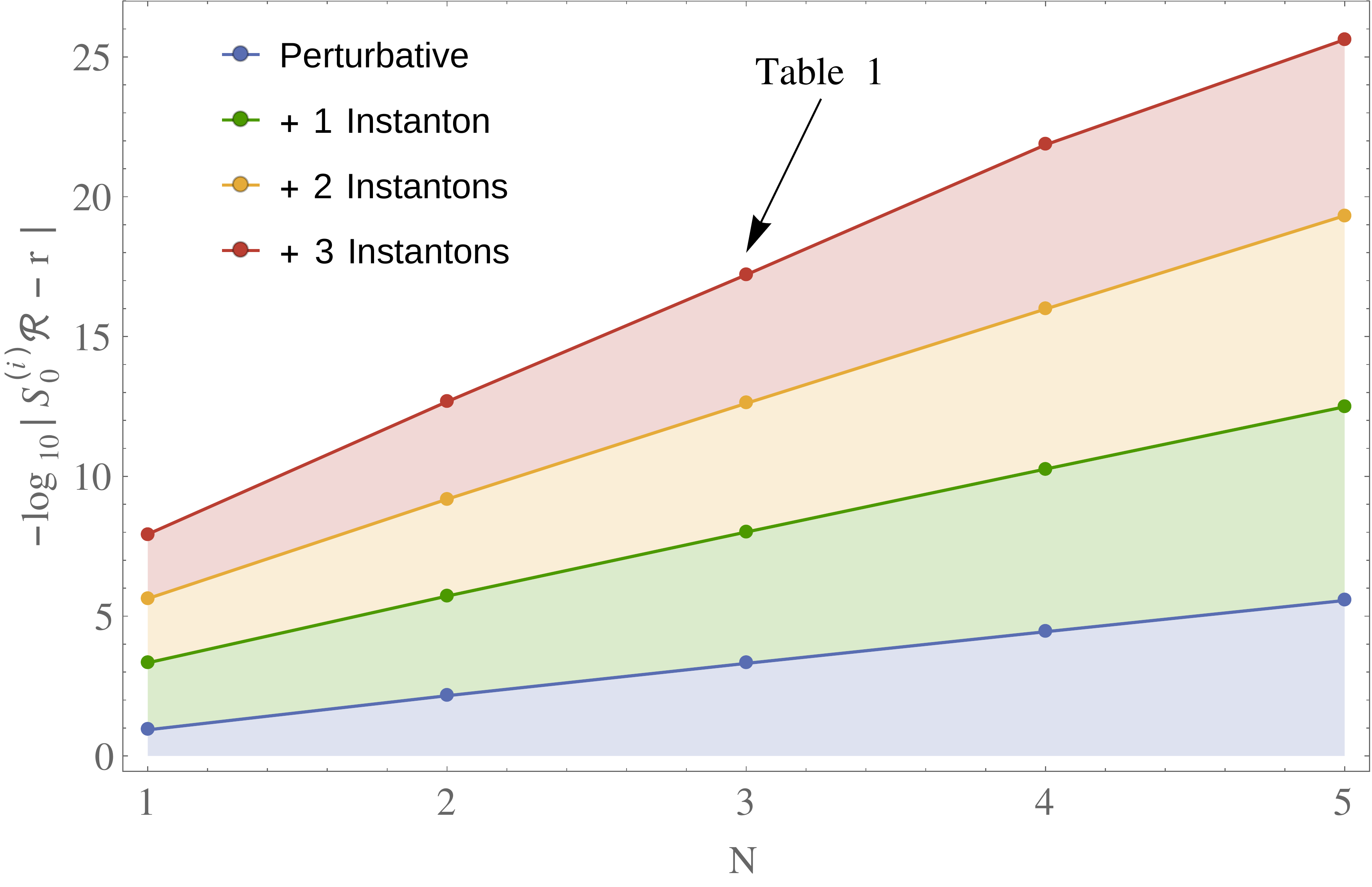}
$\quad$
\includegraphics[width=7.75cm]{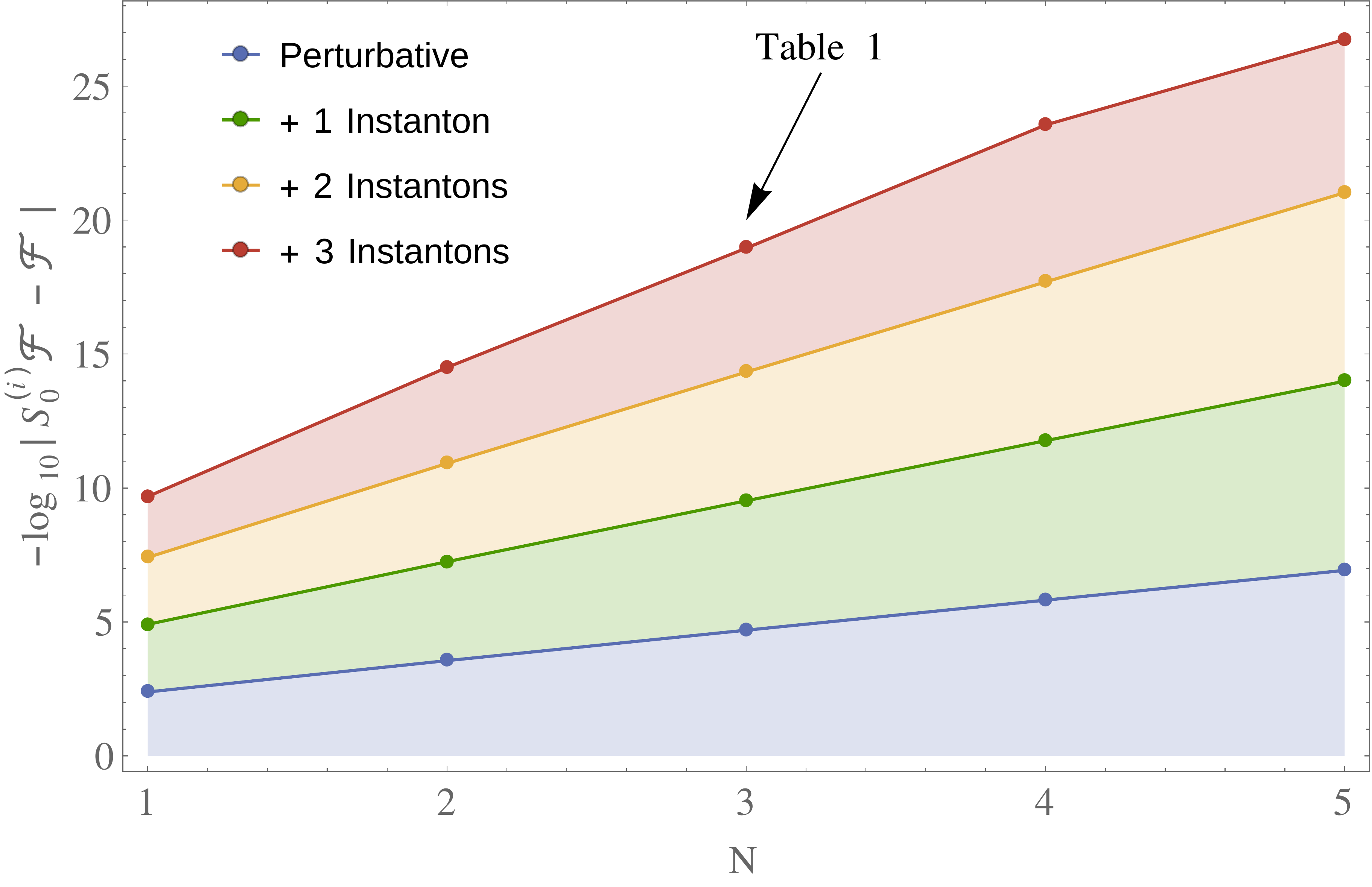}
\end{center}
\caption{Number of decimal places up to which the resummation of the recursion coefficients (left) and the free energy (right) match their exact counterparts, with $N=1,\dots,5$ and $t=6$.}
\label{fig:prec_R_F}
\end{figure}

We illustrate this visually in figure \ref{fig:prec_R_F}, also for different values of the rank. On the $x$-axis we vary $N$ and on the $y$-axis we plot $-\log_{10} \left| S^{(i)}_0 \CR - r_N \right|$ (left) and $-\log_{10} \left| S^{(i)}_0 \CF-\CF(N) \right|$ (right), where $r_N$ are the exact recursion coefficients \eqref{r_1}-\eqref{r_4} and $\CF(N)$ are the logarithms of the exact (normalized) partition functions \eqref{Z_2}-\eqref{Z_5}. This quantity effectively tells us the number of decimal places to which the analytical result and the resummation agree, and as we saw in the example above we get closer and closer to the full answer as we add more and more instanton sectors. One can check that at $N=3$ the matched digits are the ones shown in table \ref{tab:transseriesresum}. Note that we have chosen a relatively high value of $t$ where the picture is clearer. As we move to lower $t$ the instanton contributions get smaller in absolute value, and it may happen that if $S^{(0)}_0$ has not stabilized at enough digits it becomes harder to see their effect.

\begin{figure}[t!]
\begin{center}
\includegraphics[width=14cm]{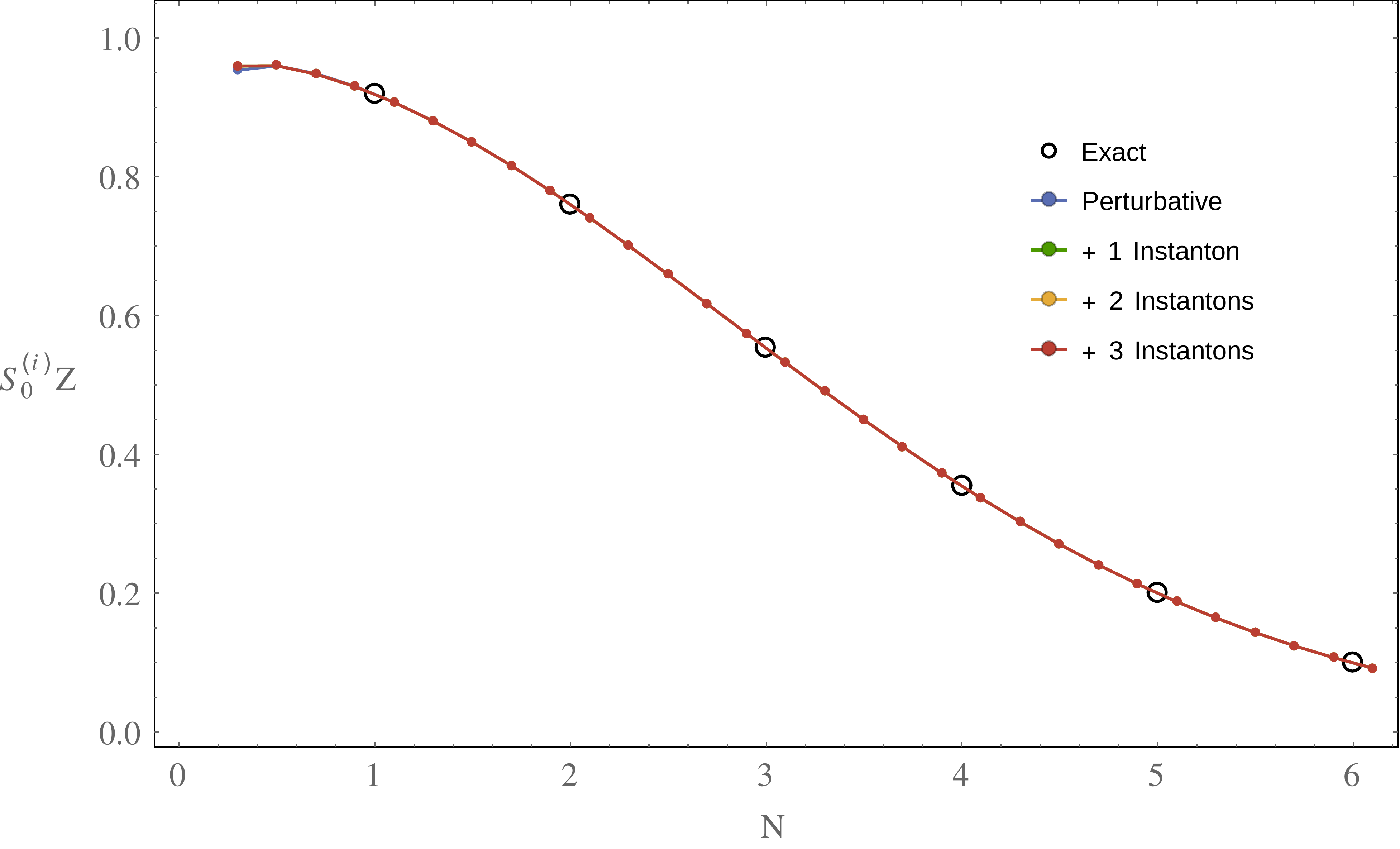}
\end{center}
\caption{Resummation of the partition function for \textit{continuous} $N \in (0,6)$, $t=1$, including up to $3$-instantons (the lines are all superposed), and the exact results at integer values of $N$.}
\label{fig:interp_Z}
\end{figure}

Having seen how the transseries so precisely captures the exact results, at small integer values of the rank, one may ask if it can go beyond this requirement and actually compute the free energy or partition function at \textit{continuous} values of $N$. It should be clear that nothing changes as one considers the resummation for non-integer $N$. Furthermore, as we have analytical results for the exact partition function at integer values of $N$, we can ask how the resummation interpolates between them. This is shown in figure \ref{fig:interp_Z}, with $t=1$, where it is clear that the resummation produces a smooth interpolation between the analytical results arising from the matrix integral. Of course these results are already going well beyond the exact matrix integral results, as the latter are not even defined when $N$ is non-integer.

It is worth pointing out that we are not resumming $Z$ in the same way we did for $\CF$ or $\CR$, since it is very inefficient to exponentiate the free energy transseries and then extract the $1/N$ coefficients at each order. Instead, we first resum the free energy and only then exponentiate the result. Furthermore, note that for the almost entirety of the plot in figure \ref{fig:interp_Z}, the resummations including any or all instanton sectors are indistinguishable and the four lines are coincident. As we showed above, these distinctions only appear after a certain number of decimal places, and this is impossible to spot in these scales. However, in the next section we will also explore complex-valued $t$ and we will see cases where there is a ``macroscopic'' distance between different resummations. Finally, we notice that the coincident lines seem to split apart as we get closer to $N=0$. The fact that we are dealing with a normalized partition function means that $Z(0)=1$. However, this is a point of infinitely strong coupling, and even the large amount of data we have is insufficient, from a numerical standpoint, to get consistent results at this point (\textit{i.e.}, technically, at infinitely strong coupling we would need infinite terms in the Borel--\'Ecalle transseries resummation, which is unachievable).

\section{Analytic Continuations and Stokes Phenomenon}\label{sec:resurgent_stokes}

In the previous section we have limited our attention to the case where $t \in \BR^+$. While definitely crediting the power of resurgent transseries in achieving to go beyond integer rank and actually \textit{define} the gauge theory partition function at continuous values of $N$, the reader might get the wrong impression that other than that all the transseries is implementing is smaller and smaller instanton corrections to the perturbative expansion, piling on top of each other. This is certainly not the generic case and such picture will dramatically change as we consider the analytic continuation onto complex values of the 't~Hooft parameter, due to Stokes phenomenon.

\begin{figure}[t!]
\begin{tabular}[b]{ccc}
\begin{subfigure}[b]{5.5cm}
\raisebox{0.75cm}{\includegraphics[width=5.5cm]{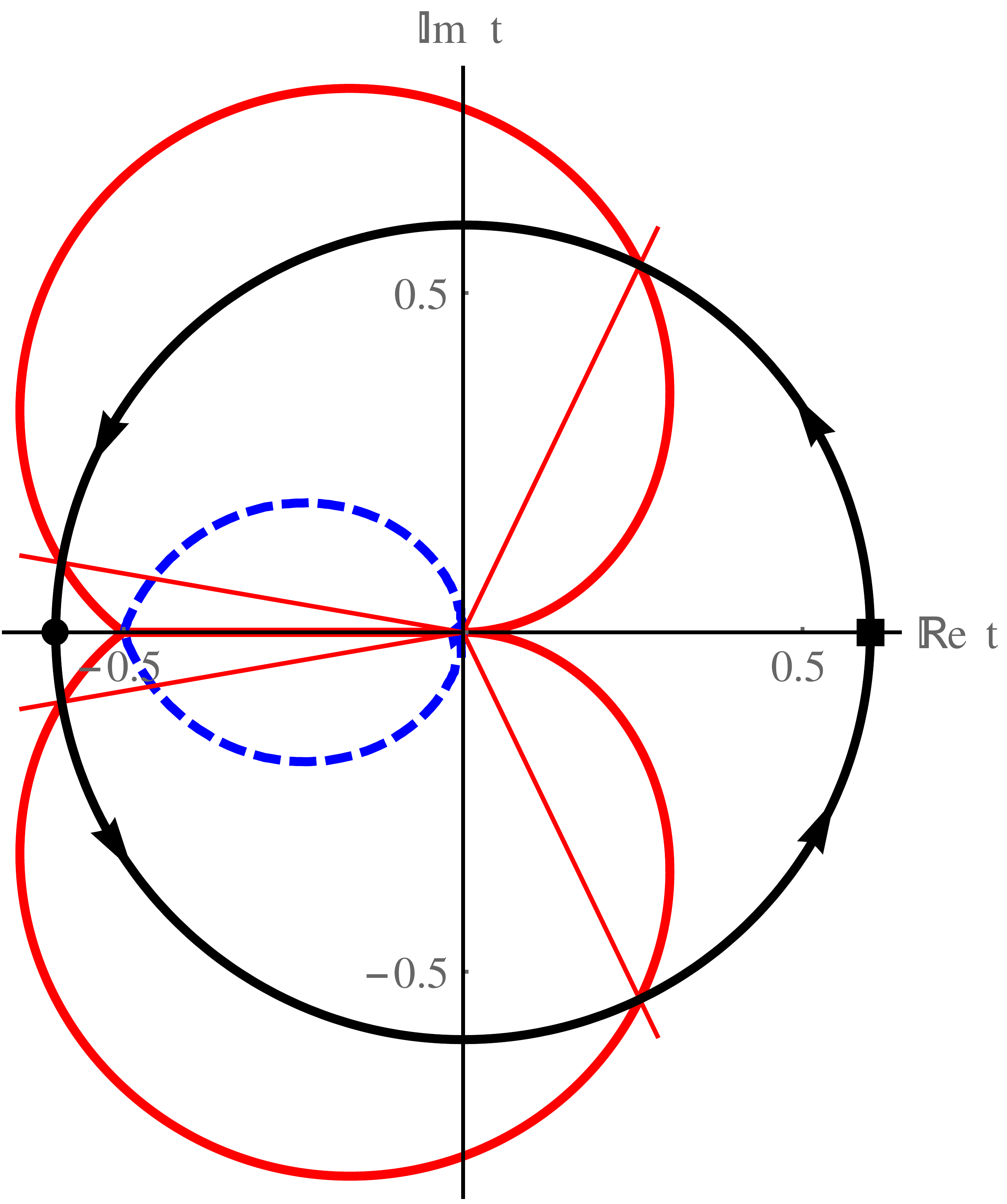}}
\end{subfigure}
&
\begin{subfigure}[b]{5.5cm}
\includegraphics[width=5.5cm]{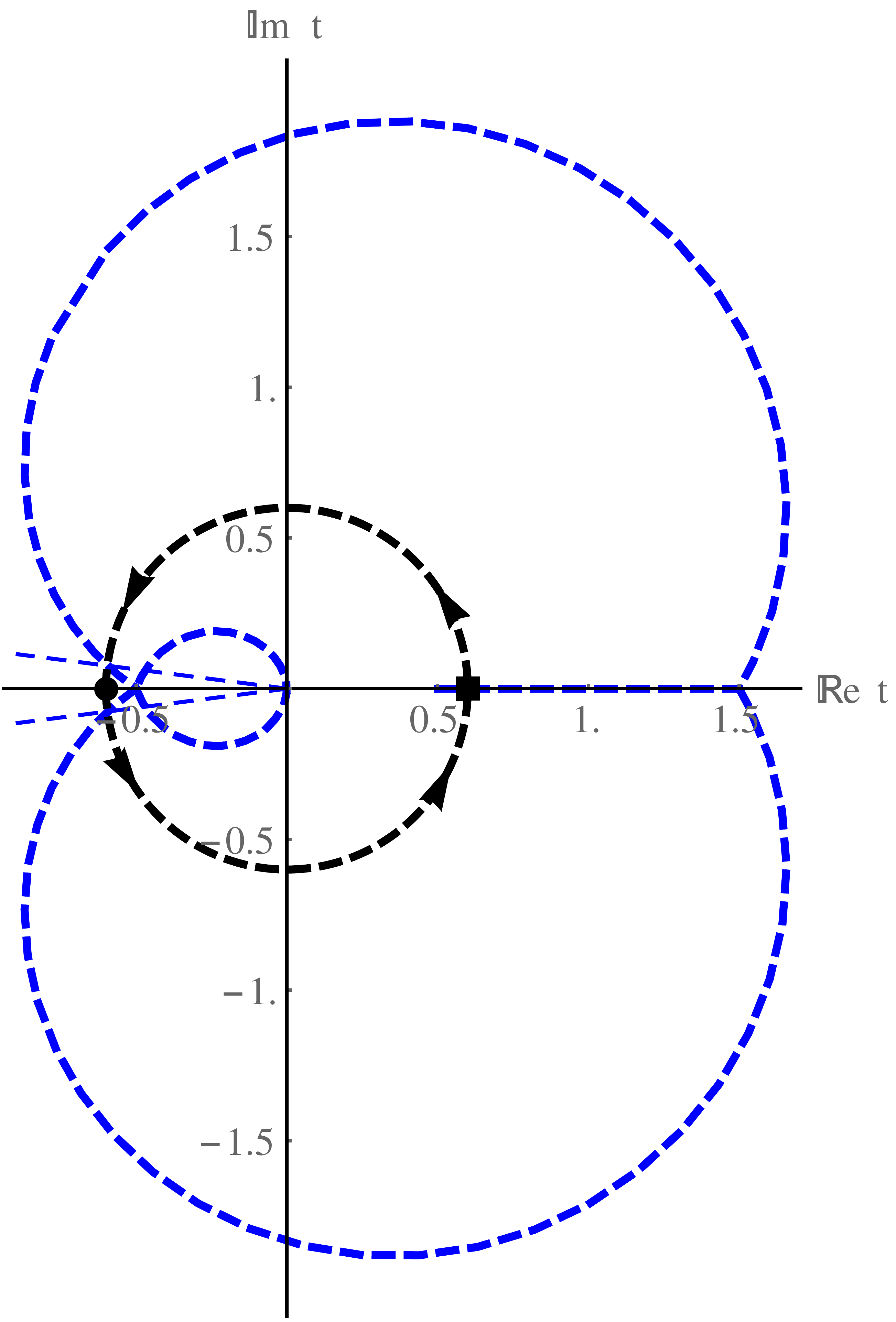}
\end{subfigure}
&
\hspace{-.3cm}\raisebox{2.5cm}{\begin{tabular}[b]{c}
\begin{subfigure}[b]{5.2cm}
\includegraphics[width=4cm]{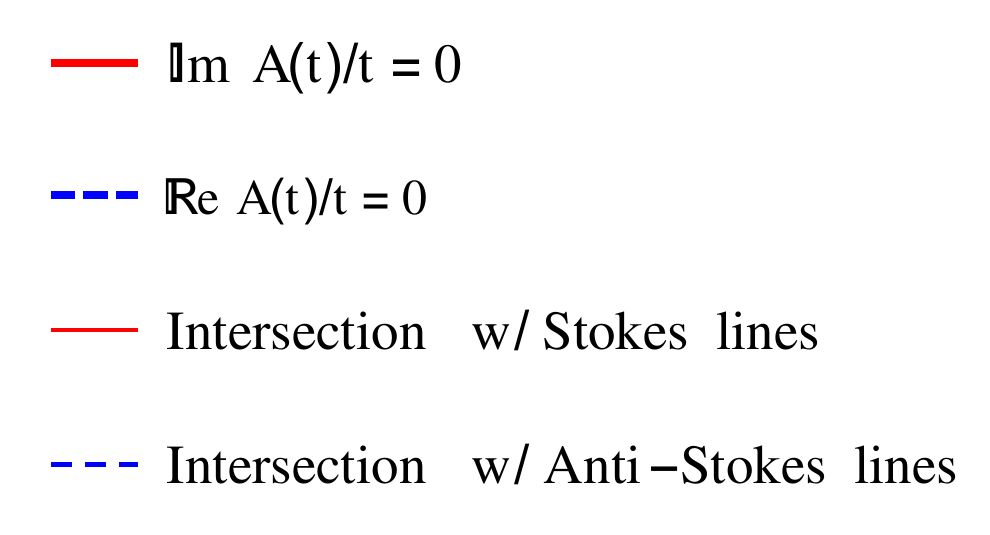}
\end{subfigure}\\
\begin{subfigure}[b]{5.2cm}
\hspace{-.5cm}\includegraphics[width=4.5cm]{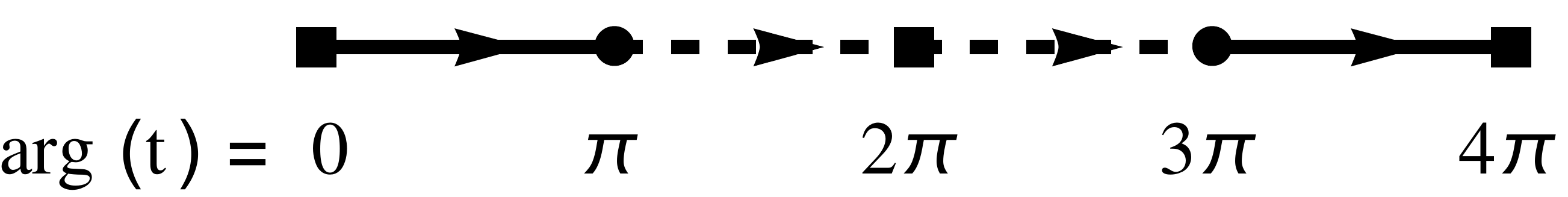}
\end{subfigure}
\end{tabular}}
\end{tabular}
\caption{Phase diagrams showing anti-Stokes (thick, dashed, blue curved) and Stokes (thick, solid, red curved) boundaries along with the path $|t|=0.6$ (arrowed black) and the intersections of this path with Stokes (thin, solid, red) and anti-Stokes (thin, dashed, blue) lines. The left plot represents $\arg(t) \in ( -\pi,+\pi)$ while the one on the right is for $\arg(t) \in ( \pi,3\pi)$. In the last line of the caption we show how the motion on the two sheets takes place as $\arg(t)$ is changed from $0$ (the solid square on the first diagram) all the way (back) to $4\pi$. Note that there is a third intersection with an anti-Stokes line in the second sheet, for $\arg(t)=2\pi$, but we do not draw the intersecting ray because it would lie along the positive real axis.}
\label{fig:phasediagramv1}
\end{figure}

\begin{figure}[ht!]
\begin{center}
\includegraphics[width=10.3cm]{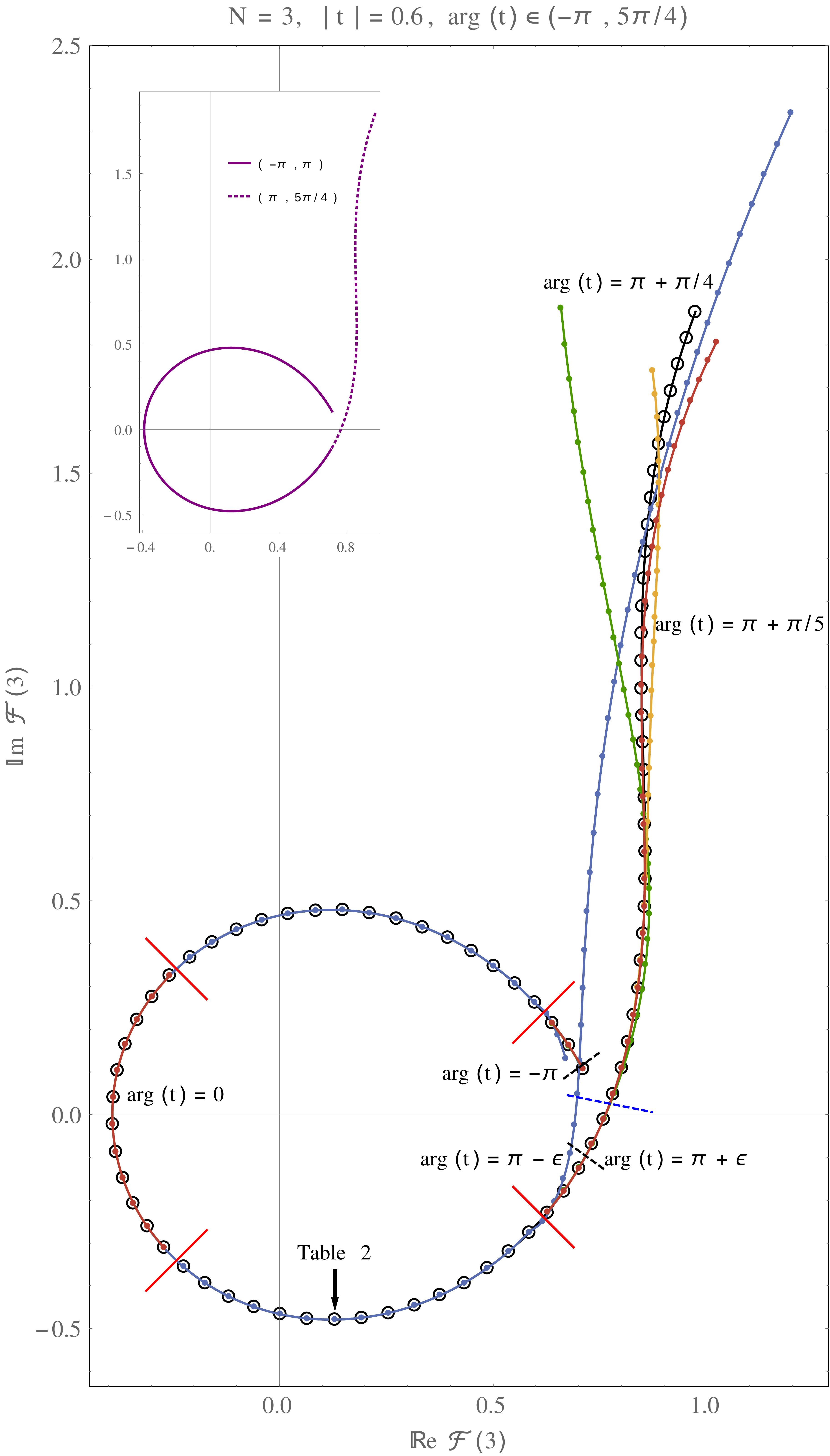}
\quad
\raisebox{7.5cm}{\includegraphics[width=3.5cm]{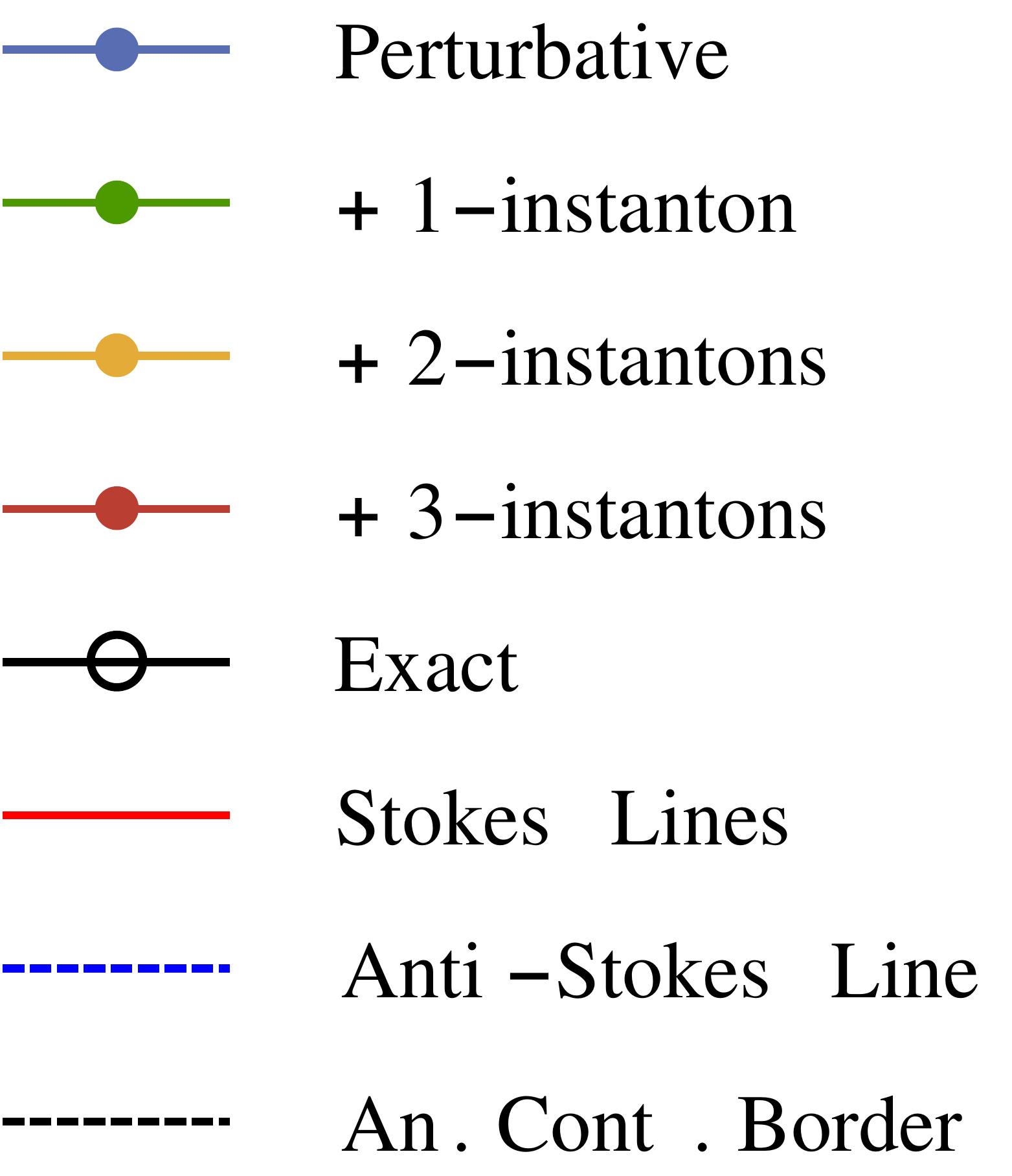}} 
\end{center}
\vspace{-1\baselineskip}
\caption{Resummation of the free energy transseries for $|t|=0.6$ and $N=3$, taking into account the different values of $\sigma$ (due to Stokes phenomenon) depending on $\arg(t) \in (-\pi, \pi + \pi/4)$. The Stokes lines illustrate how in some regions $\sigma=0$ and the perturbative resummation yields the correct result, while in other regions $\sigma=S_1$ and we need to include up to three instantons in order to reproduce the exact result (shown in the left plot of figure \ref{fig:F23_monodromy} and enclosed here). It is also clear how past the anti-Stokes phase boundary the perturbative contribution is no longer reliable and instantons are required to take over in order to yield the correct monodromy results.}
\label{fig:resumberryplotFex1}
\end{figure}

\begin{figure}[ht!]
\begin{center} 
\includegraphics[width=12.5cm]{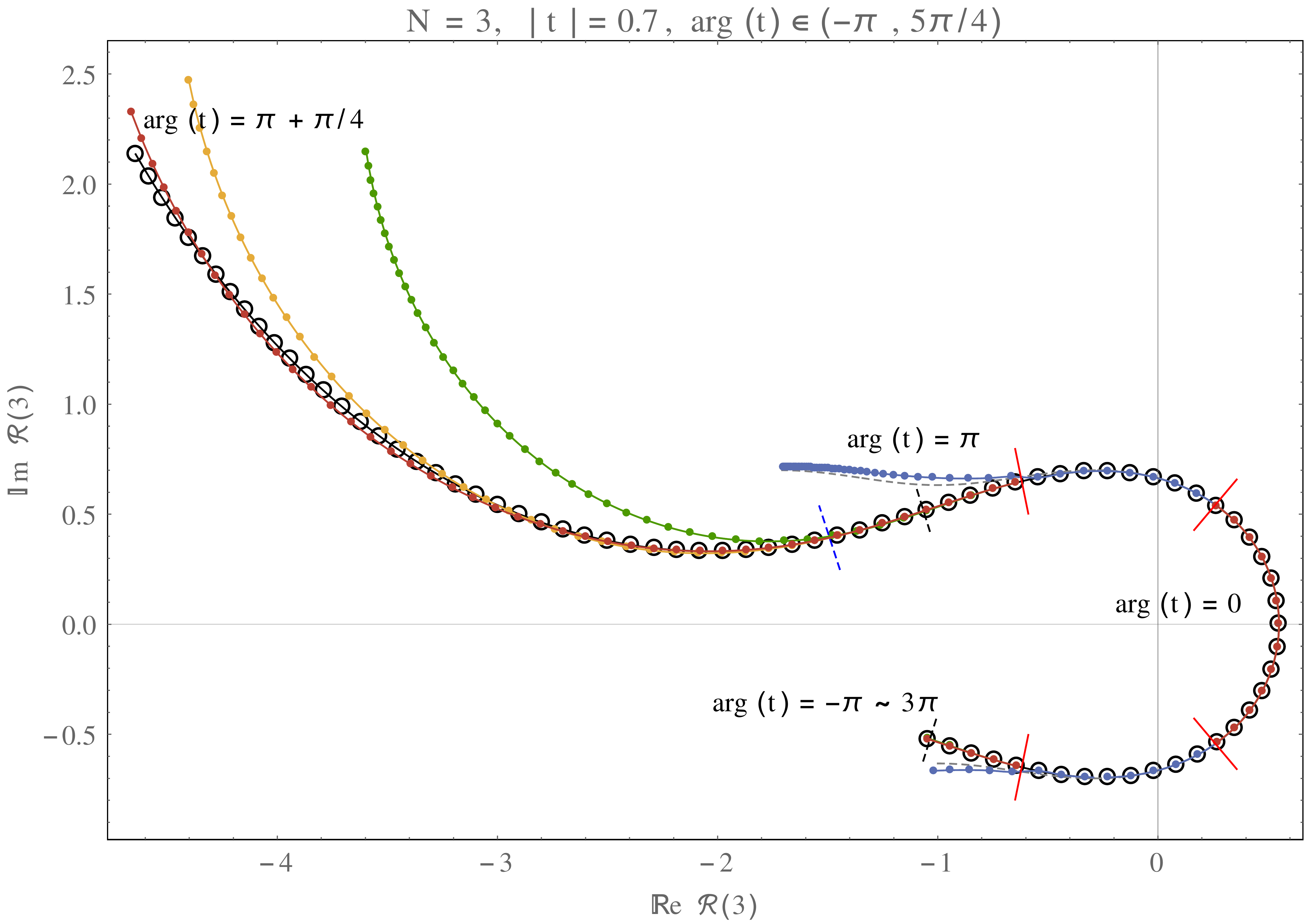}
\,
\raisebox{2cm}{\includegraphics[width=3.25cm]{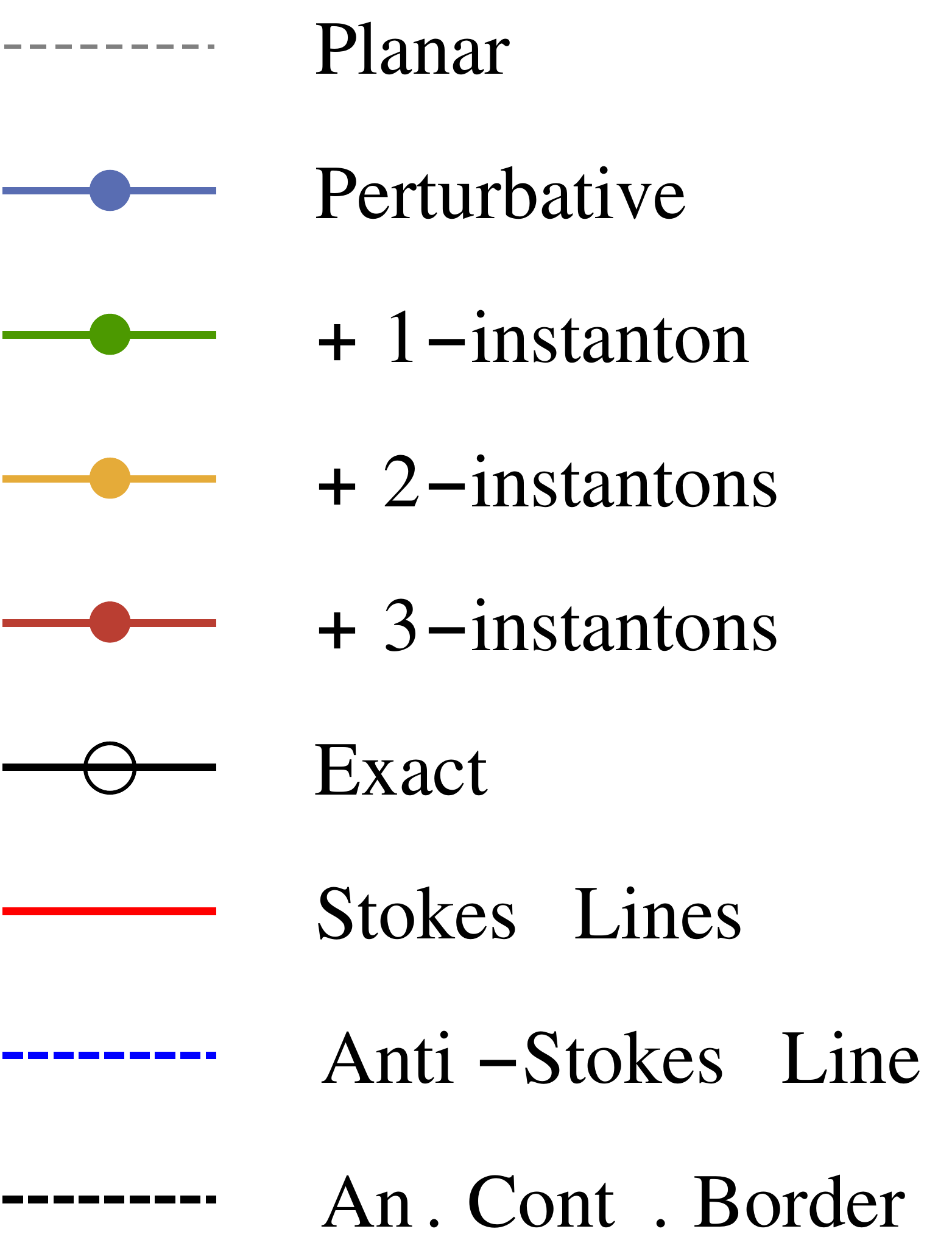}}
\end{center}
\vspace{-1\baselineskip}
\caption{Resummation of $r_3$, for $|t|=0.7$, in direct correspondence with the left  plot in figure \ref{fig:r3_monodromy}. Note the various Stokes lines in the principal domain for $\arg(t)$, and the anti-Stokes phase boundary in the second domain indicating that instanton corrections have become of order one.}
\label{fig:resumberryplotRex1}
\end{figure}

\begin{figure}[ht!]
\begin{center} 
\includegraphics[width=12.75cm]{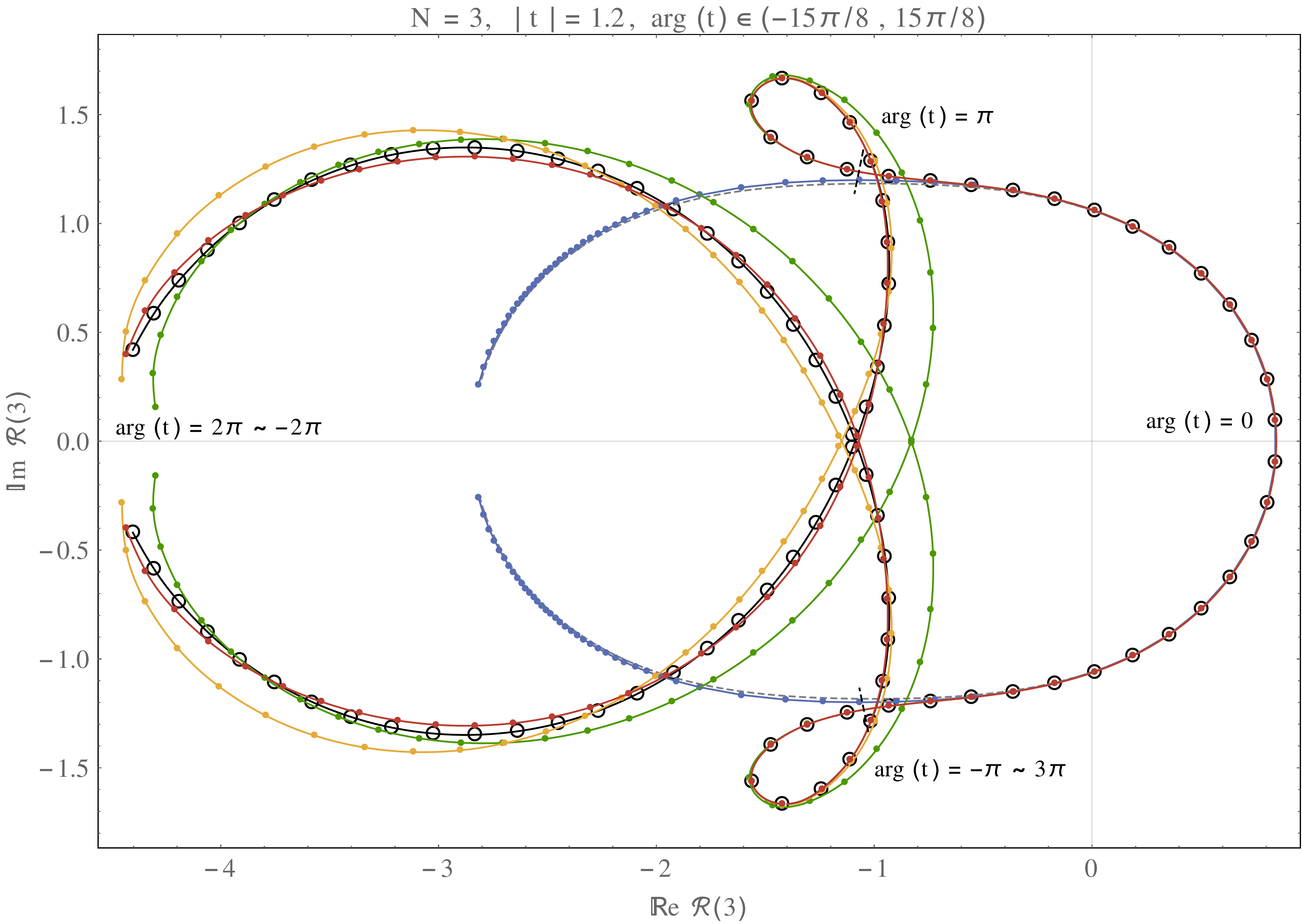}
\raisebox{3.5cm}{\includegraphics[width=3.2cm]{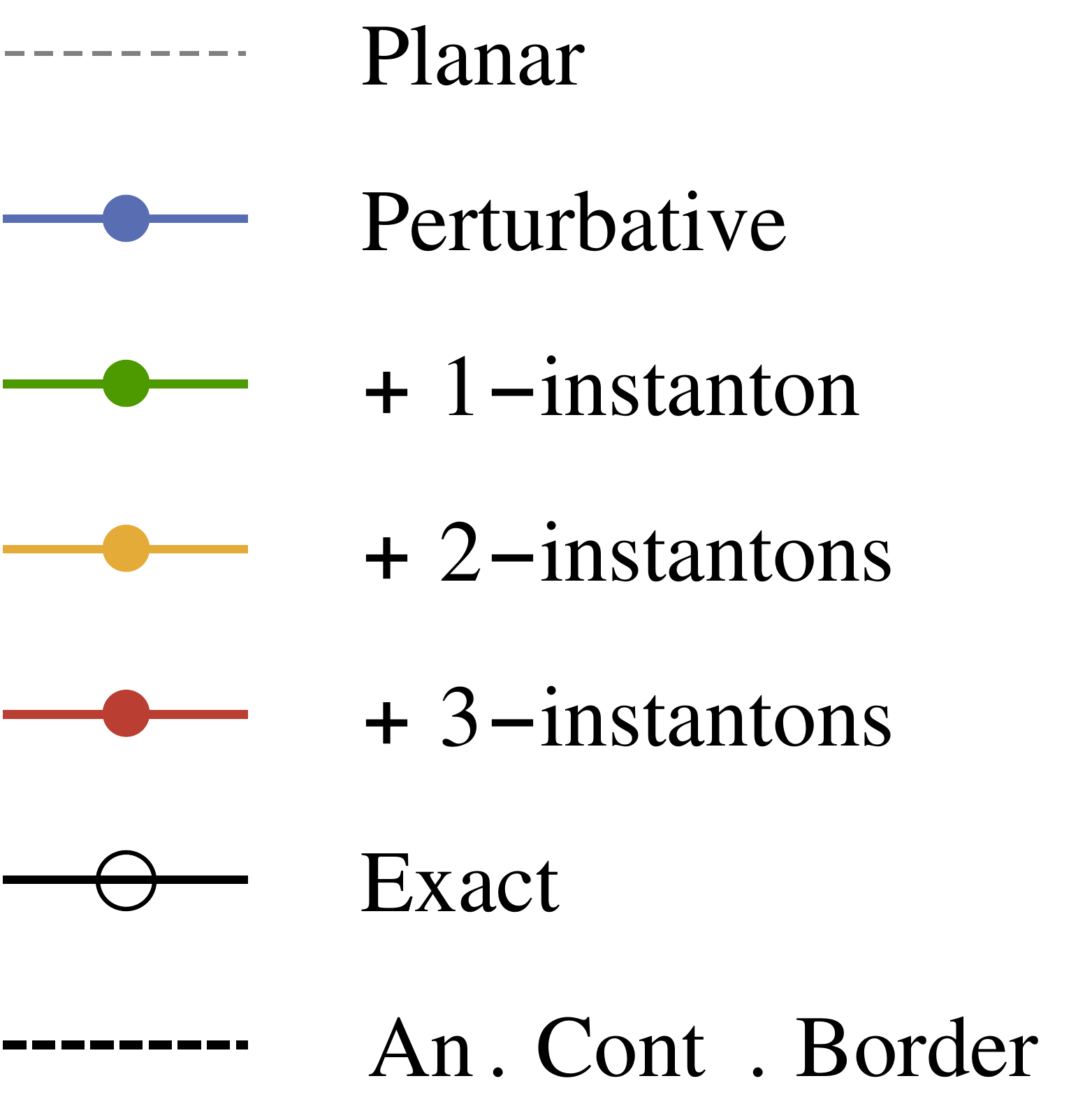}}
\end{center}
\vspace{-1\baselineskip}
\caption{Resummation of $r_3$, for $|t|=1.2$, in direct correspondence with the top right plot in figure \ref{fig:r3_monodromy}. This plot illustrates very clearly how the perturbative resummation keeps following the planar approximation (and thus producing an incorrect result), while instanton effects take dominance in order to produce the correct monodromy results.}
\label{fig:resumberryplotRex2}
\end{figure}

\begin{figure}[ht!]
\begin{center} 
\includegraphics[width=12.5cm]{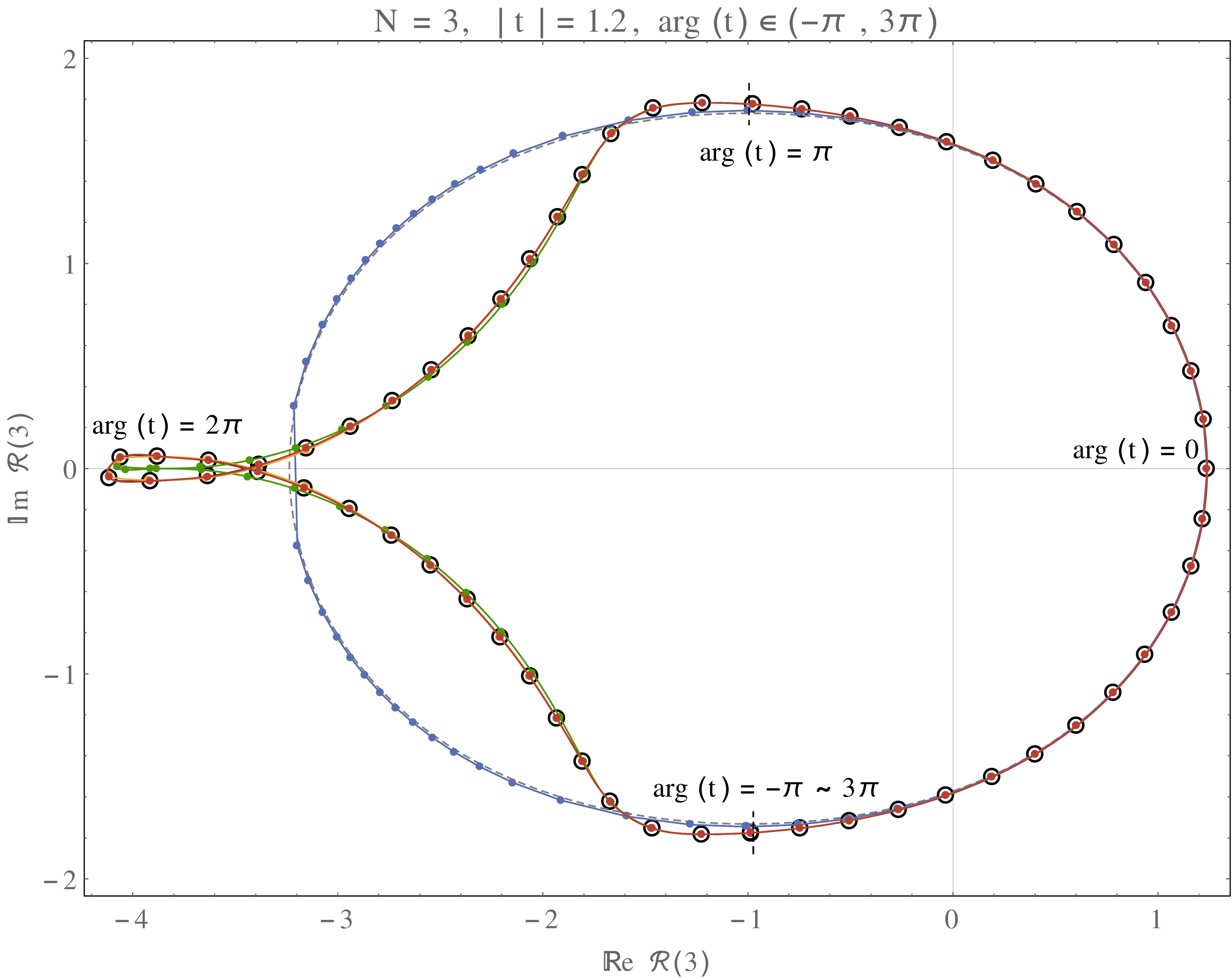}
\,
\raisebox{3.5cm}{\includegraphics[width=3.25cm]{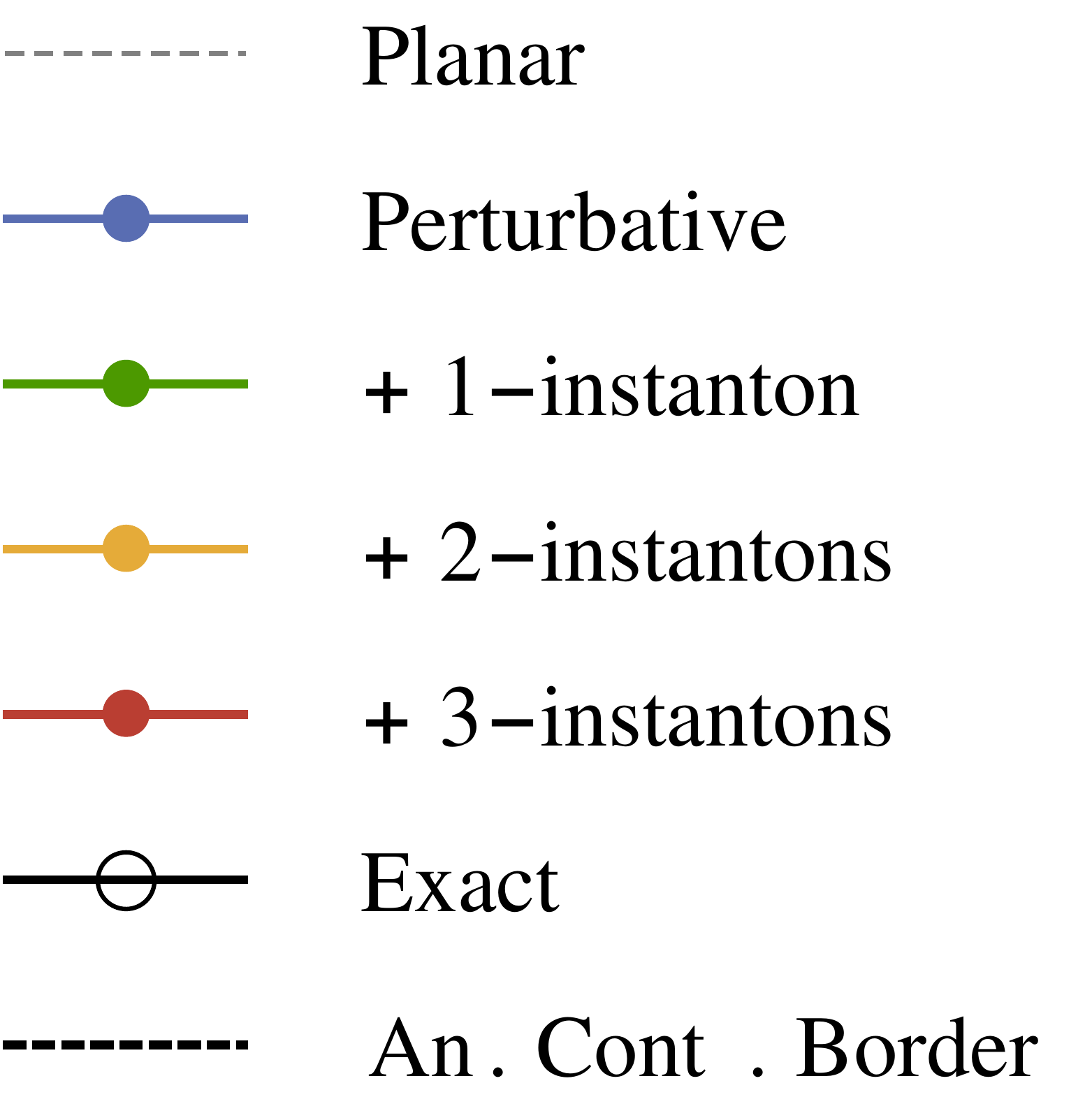}}
\end{center}
\vspace{-1\baselineskip}
\caption{Resummation of $r_3$, for $|t|=2$, in direct correspondence with the bottom right plot in figure \ref{fig:r3_monodromy}. Note how the perturbative resummation keeps following the planar approximation.}
\label{fig:resumberryplotRex3}
\end{figure}

\begin{figure}[ht!]
\begin{center} 
\includegraphics[width=13cm]{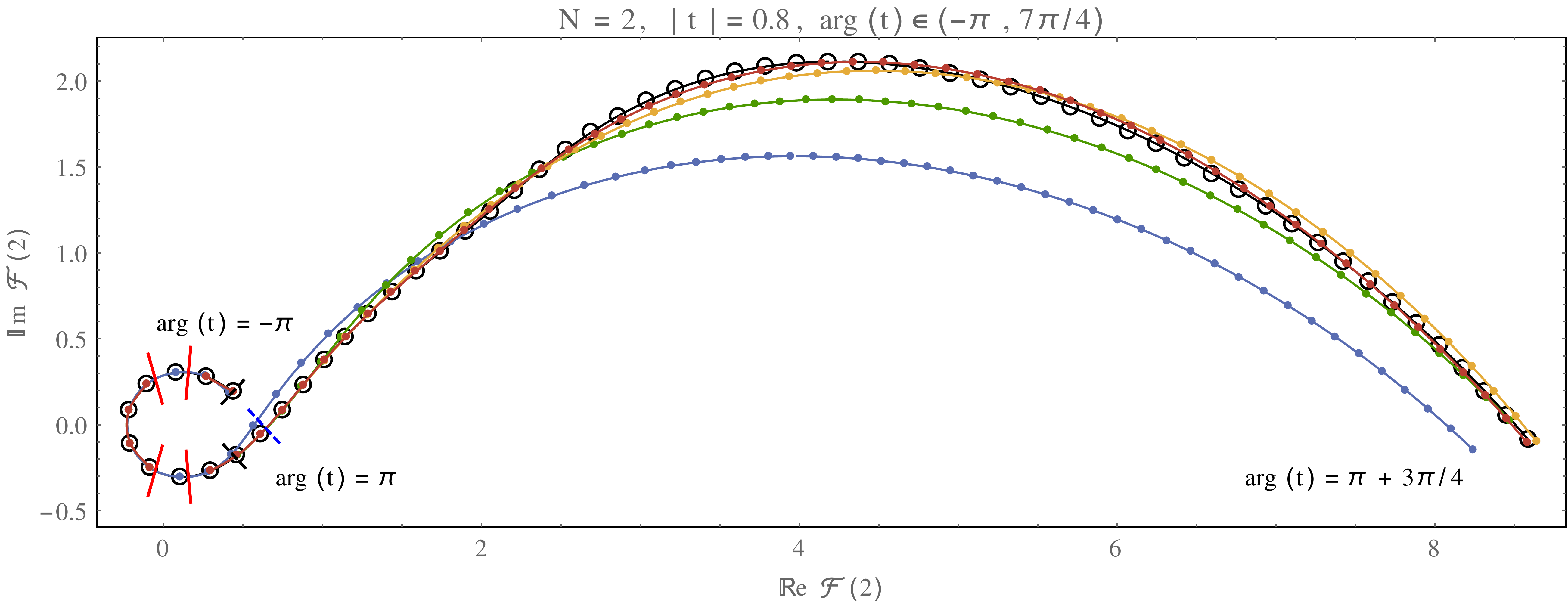}
\,
\raisebox{1cm}{\includegraphics[width=2.75cm]{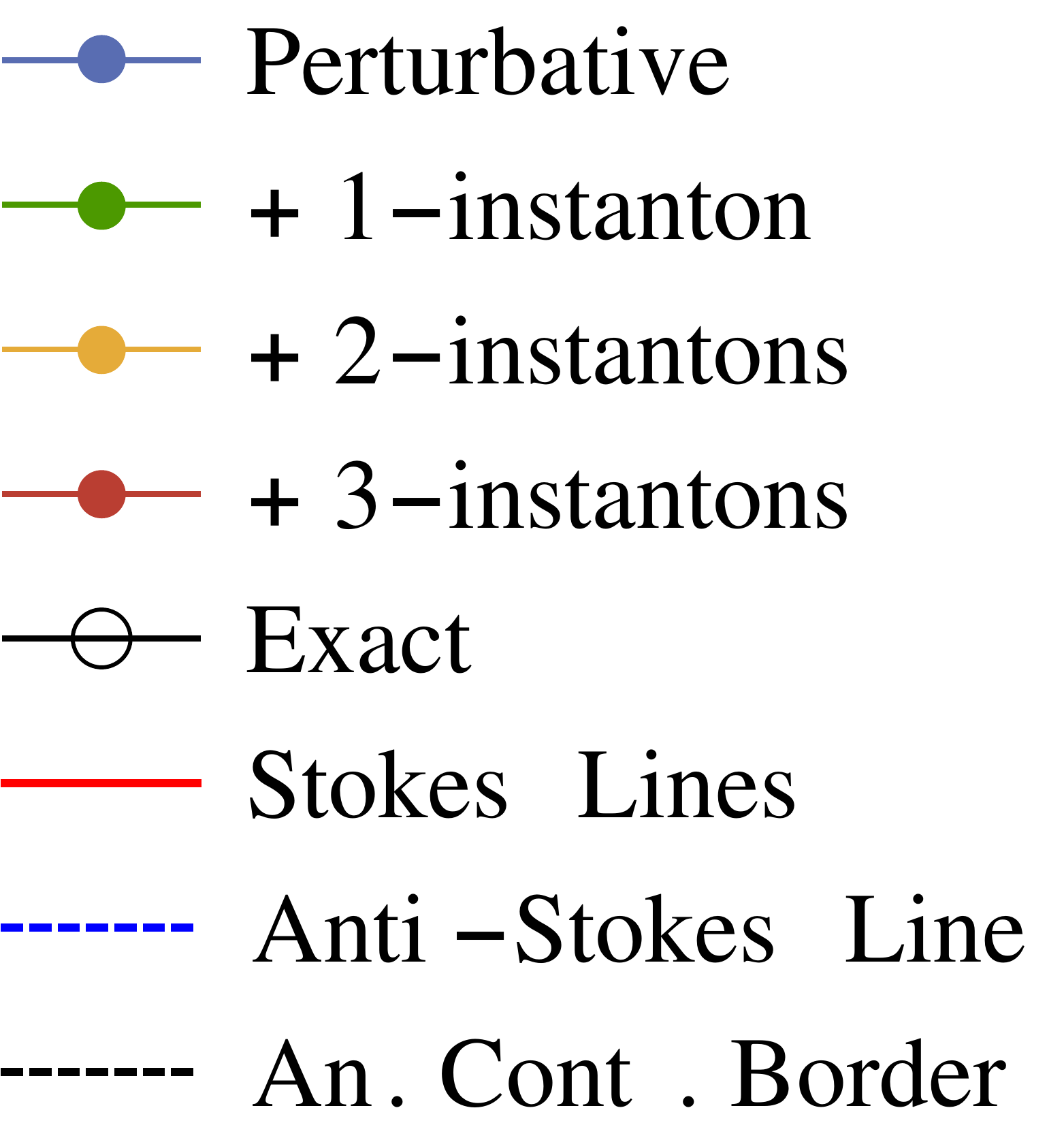}}
\end{center}
\vspace{-1\baselineskip}
\caption{Resummation of $\CF(2)$, for $|t|=0.8$, in direct correspondence with the top right plot in figure \ref{fig:F23_monodromy}. While the perturbative resummation still follows the general trend of the monodromy, it is no longer reliable past the anti-Stokes phase boundary.}
\label{fig:resumberryplotFex3}
\end{figure}

\begin{figure}[ht!]
\begin{center} 
\includegraphics[width=10cm]{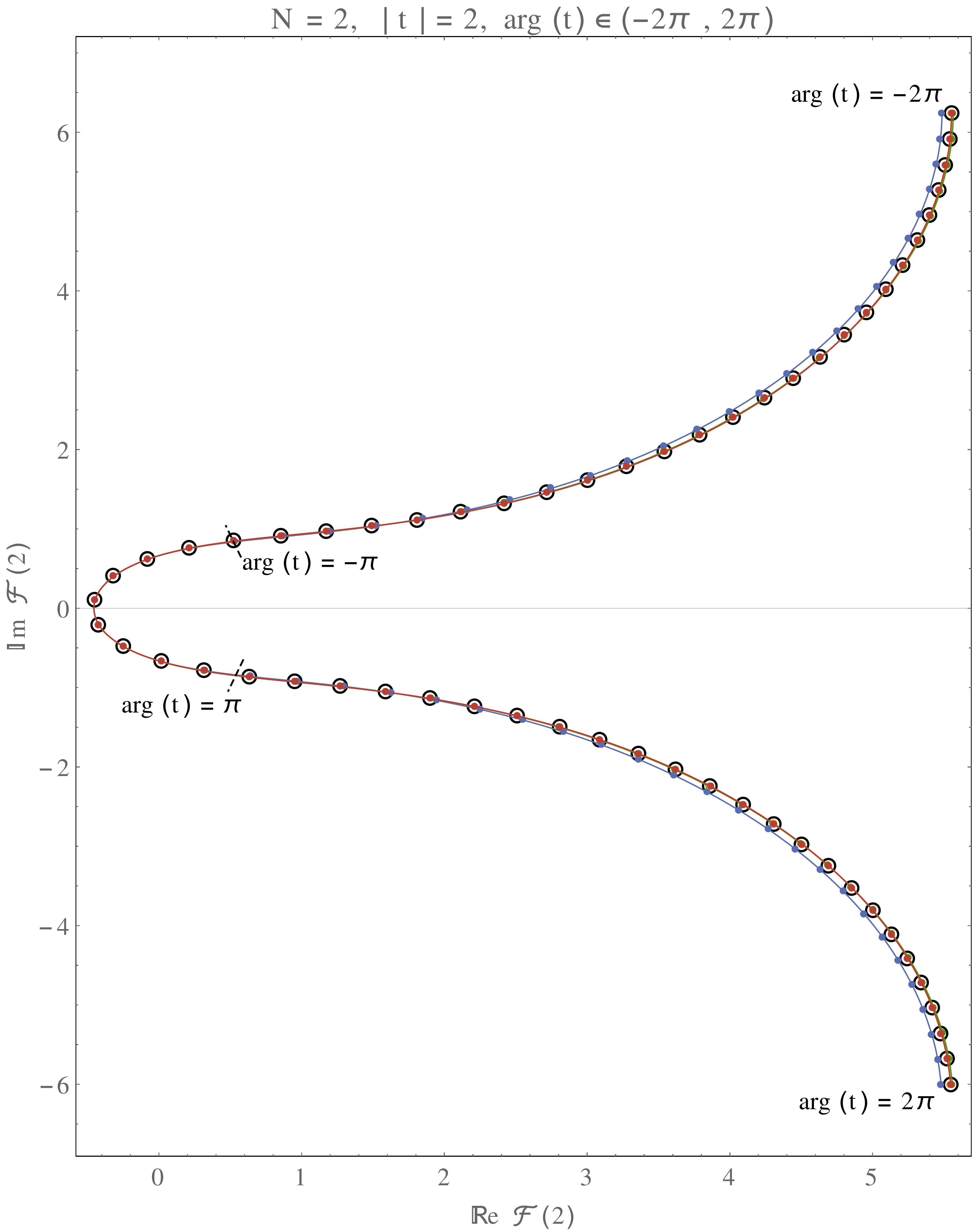}
\,
\raisebox{5cm}{\includegraphics[width=3cm]{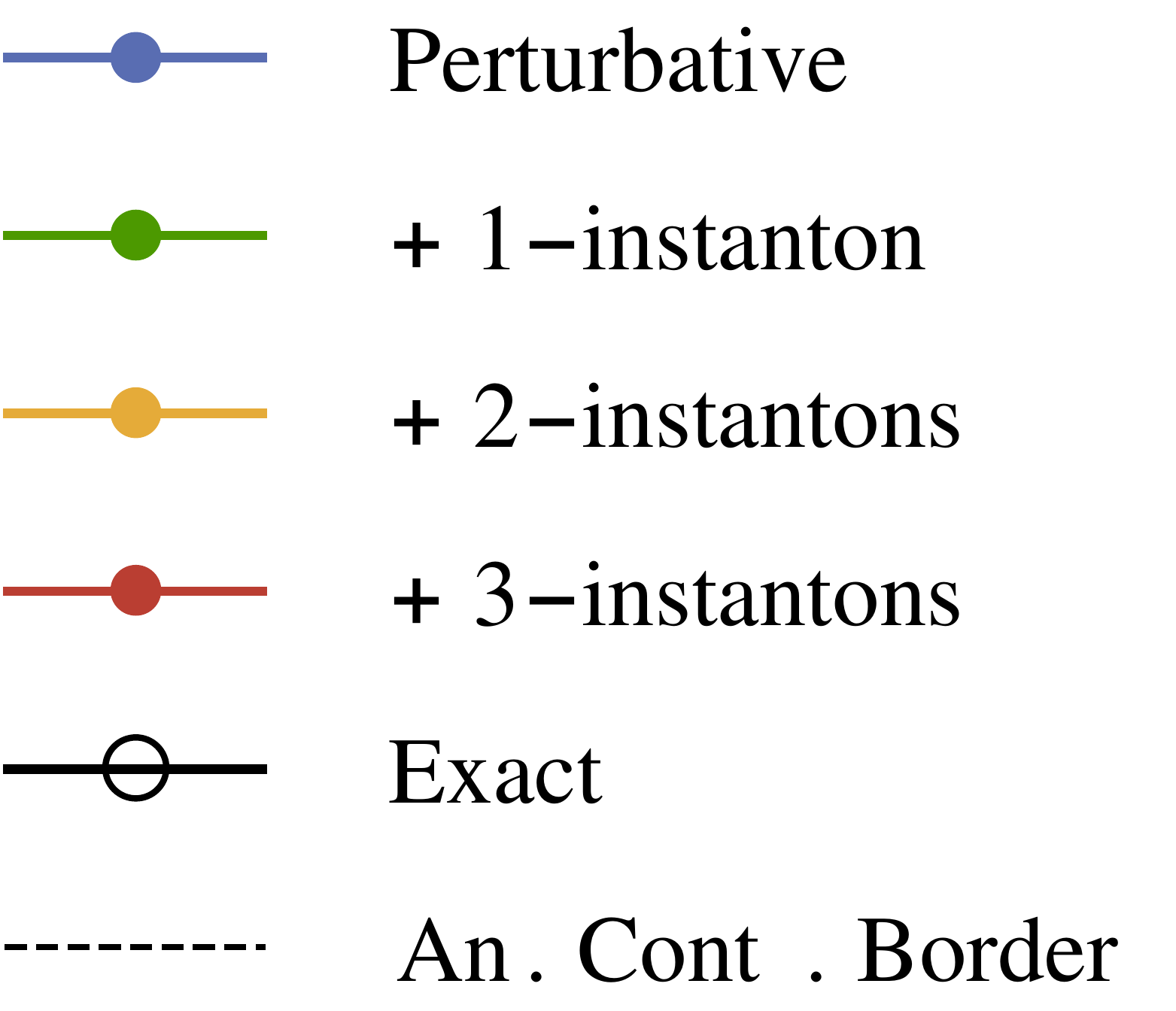}}
\end{center}
\vspace{-1\baselineskip}
\caption{Resummation of $\CF(2)$, for $|t|=2$, in direct correspondence with the bottom right plot in figure \ref{fig:F23_monodromy}. Unlike previous plots, note how here the perturbative result is pretty reliable.}
\label{fig:resumberryplotFex2}
\end{figure}

\begin{figure}[ht!]
\begin{center}
\hspace{-1cm}
\includegraphics[width=14cm]{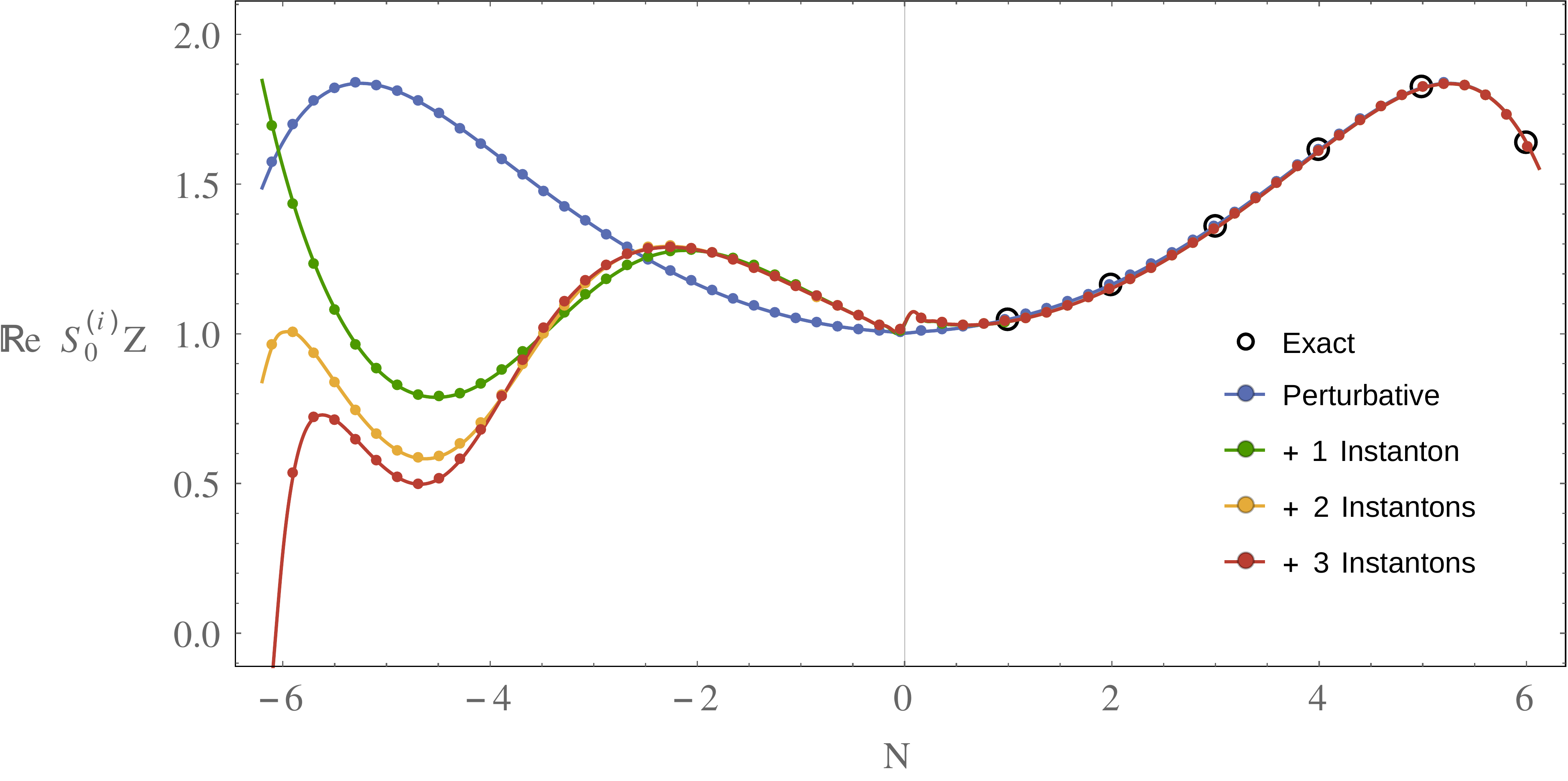}
\end{center}
\caption{Real part of the resummed partition function, $S_0 Z$, for continuous negative to positive $N \in (-6,6)$ and fixed $t = \frac{1}{2}\, \rme^{\frac{5\pi\rmi}{6}}$. The  accuracy is no longer reliable for $N \lesssim -4$.}
\label{fig:negative_N}
\end{figure}

\begin{figure}[ht!]
\begin{center}
\includegraphics[width=14cm]{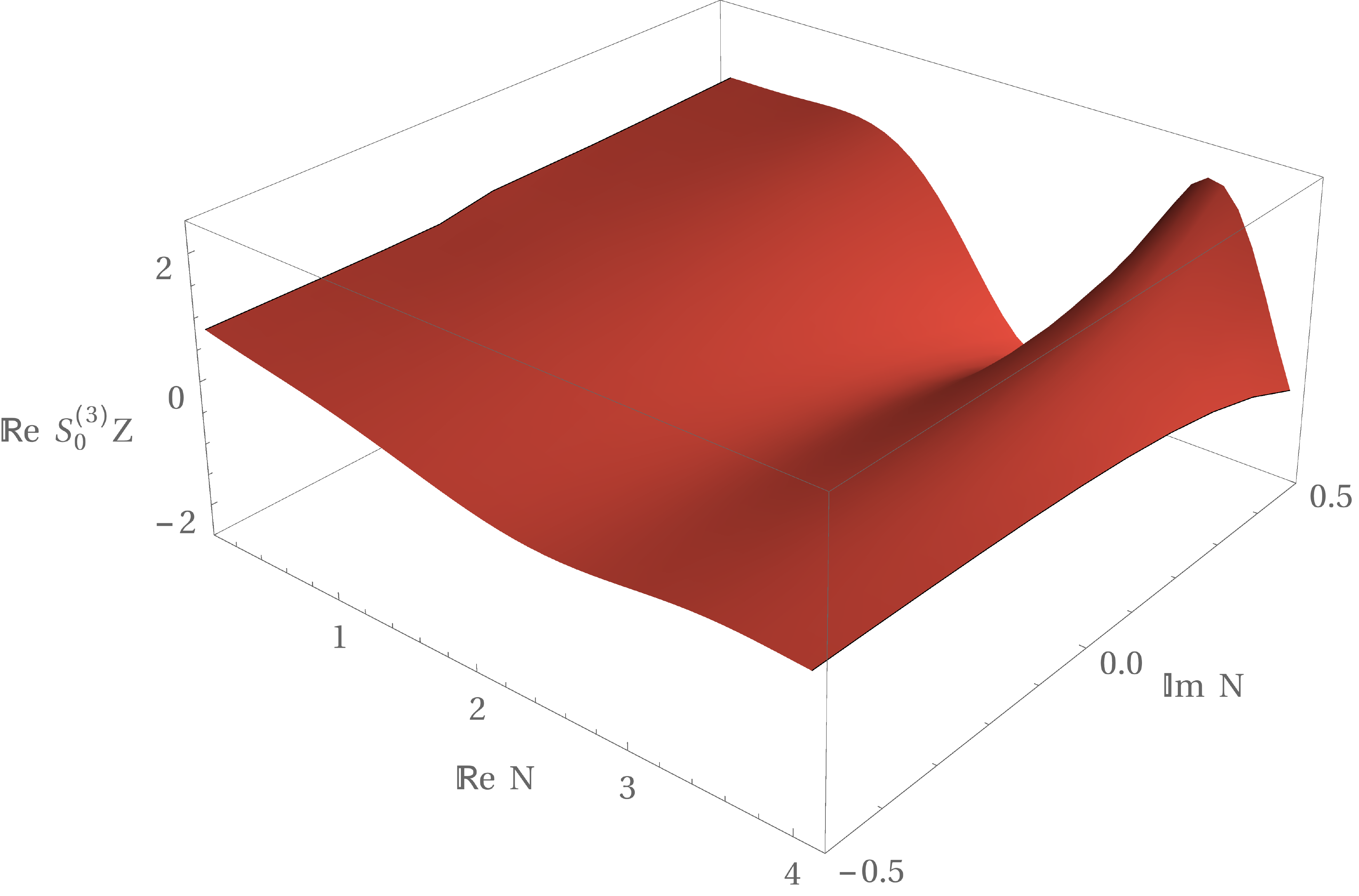}
\end{center}
\caption{Real part of the partition function, resummed up to three instantons, $S^{(3)}_0 Z$, for \textit{complex} rank $\re(N) \in (0,4)$ and $\im(N) \in (-1/2, 1/2)$, and fixed  $t = 10\, \rme^{\frac{99\pi\rmi}{100}}$.}
\label{fig:complex_N}
\end{figure}

To understand how this works, let us first address an illustrative example. Consider the free energy when $N=3$ and at fixed 't~Hooft coupling $|t| = 0.6$ but varying argument $\arg(t) \in (-\pi, \pi+\pi/4 )$. This is a continuous function with nontrivial monodromy, as shown in figure~\ref{fig:F23_monodromy}, computed from analytic expressions. In the 't~Hooft limit, the large $N$ resurgent transseries for $\CF$, \eqref{Ftransseries}, must have all the required information in order to reproduce this plot. To extract it we have to resum it, as explained in the previous section, but also take into account Stokes phenomenon. This last step is crucial and in practice it implies selecting a particular member from the family of transseries parametrized by $\sigma$. For $t\in\BR^+$ we saw that the correct value is given by \eqref{valuesigma}, but as we now vary $\arg(t)$  we will need to implement Stokes transitions as shown in \eqref{Stokes}. This is a general feature of representing functions as transseries: we need the whole family of solutions, as parametrized by the  parameter $\sigma$, and a practical understanding of how Stokes phenomenon selects the right member as we move across the complex $t$-plane. 

Stokes transitions take place at Stokes lines, the rays where multi-instanton singularities lie on the complex Borel plane. In the complex 't~Hooft plane they satisfy $\im \left(A(t)/t\right) = 0$, and impose the familiar jump in the transseries parameter. One may also cross anti-Stokes lines, where $\re \left(A(t)/t\right) = 0$, with instantons taking dominance over the perturbative series---in fact at this point all contributions, perturbative and nonperturbative, are of the same order. As we move beyond it and into a region where $\re \left(A(t)/t\right) < 0$, the instanton contributions are no longer exponentially suppressed. However, this does not mean that the transseries representation breaks down. The transseries is a formal object that includes complex instantons in general, but where the label ``exponentially suppressed'' or ``exponentially enhanced'' only applies in the formal large $N$ limit. In the resummation process where $N$ becomes finite, even small, and where we may venture into the complex plane, this distinction is somewhat irrelevant and even not appropriate any more. The separation between perturbative and (multi) instanton sectors is only set up at the initial definition of the transseries, where one finds out that for some values of $t$ the resummation of the perturbative series alone is enough to give the full answer, thus having $\sigma = 0$. But this washes aways at finite $N$ and complex $t$ where the resummation of the transseries yields, in practice, a power-series in $\sigma$. All these features will be clear in the examples that follow.

Graphically we can represent the Stokes and anti-Stokes boundaries in the complex $t$-plane\footnote{These become the large $N$ phase diagram for the quartic matrix model; see \cite{bt11, asv15, csv15} for further discussions.} and then, for each particular case, determine the corresponding lines as intersections with the path $|t|=\text{constant}$. We show this in figure \ref{fig:phasediagramv1} for our example value $|t| = 0.6$, where the diagram on the left corresponds to the first sheet, $\arg(t)\in(-\pi,\pi)$, and the one on the right to the second sheet, $\arg(t)\in(\pi,3\pi)$. We see that in the first sheet there are four Stokes lines at different arguments $\arg(t)\in(-\pi,\pi)$, and in the second sheet there are three anti-Stokes phase boundaries. There is also a symmetry with respect to the real line, so we can focus on $\arg(t) \geq 0$. 
Of course the information shown in figure \ref{fig:phasediagramv1} is incomplete without the actual resummation of the transseries, for the different values of $\arg(t)$, and that is displayed in figure \ref{fig:resumberryplotFex1}. In the following we shall explain both these plots\footnote{A very pedagogical introduction to this type of Argand plots may be found in the excellent review \cite{b91}, addressing the simple example of the Airy function (which is a linear problem with no multi-instanton sectors). A similar example but for the Bessel function (again linear and without multi-instantons) appears in \cite{cku14}.} at the same time, moving along the arrowed path drawn on the phase diagram, figure \ref{fig:phasediagramv1}, and then looking at the relevant features of the resummed transseries, figure \ref{fig:resumberryplotFex1}.

For $\arg(t) = 0$ we are in a case similar to the one described in the previous section: we can reproduce the value of the free energy with increasing accuracy by piling up smaller and smaller instanton contributions on top of the (already quite accurate) perturbative result. The value of the transseries parameter is \eqref{valuesigma}, \textit{i.e.}, the Stokes constant. The reader may also want to take another look at table \ref{tab:transseriesresum} and figure \ref{fig:prec_R_F} from the previous section\footnote{Note that in the present section we focus on a smaller value of $|t|$ than in the previous section, to illustrate Stokes phenomena. However the precision of the results is reduced, due to the finite number of available transseries coefficients. In practice, the contributions of the second and third instantons do not provide stable digits to display. In any case, in figure \ref{fig:resumberryplotFex1} the different contributions cannot be distinguished with the naked eye.}. As we increase $\arg(t)$ from $0$ onwards we will cross the first Stokes line, where a transition occurs that selects the transseries with $\sigma = 0$. That is, we reach a region where perturbation theory alone is enough to reproduce the exact value of the free energy. To show that this is the case, we display the different instanton contributions in table \ref{tab:resumjustpert}. It is clear that if we were to add them to the perturbative result we would immediately deviate from the exact result!
\begin{table}[h!]
\begin{center}
\begin{tabular}{rccc}
Sector & \hspace{0.0cm} & $\re\,S_0 \CF^{(n)}$ & $\im\,S_0 \CF^{(n)}$ \\[3pt]
\hline 
& & &\\[-8pt]
Perturbative && $+0.130\,991\,945\,237\,228\ldots$ & $-0.478\,840\,360\,187\,836\ldots$ \\
{\textcolor{orange} {\sout{{\textcolor{black}{1-Instanton}}}}} && {\textcolor{orange} {\sout{{\textcolor{black}{$-0.000\,070\,474\,759\,944\ldots$}}}}} & {\textcolor{orange} {\sout{{\textcolor{black}{$-0.000\,010\,860\,987\,563\ldots$}}}}} \\
{\textcolor{orange} {\sout{{\textcolor{black}{2-Instanton}}}}} && {\textcolor{orange} {\sout{{\textcolor{black}{$-0.000\,000\,002\,360\,007\ldots$}}}}} & {\textcolor{orange} {\sout{{\textcolor{black}{$-0.000\,000\,001\,378\,327\ldots$}}}}}\\
{\textcolor{orange} {\sout{{\textcolor{black}{3-Instanton}}}}} && {\textcolor{orange} {\sout{{\textcolor{black}{$-0.000\,000\,000\,000\,097\ldots$}}}}} & {\textcolor{orange} {\sout{{\textcolor{black}{$-0.000\,000\,000\,000\,095\ldots$}}}}} \\
& & &\\[-10pt]
Exact && $+0.130\,991\,945\,237\,228\,\ldots$ & $-0.478\,840\,360\,187\,836\ldots$
\end{tabular}
\end{center}
\caption{Comparison of the real and imaginary parts of the resummed $\CF$-transseries, at the instanton sector $n = 0,1,2,3$, and compared against the exact result for $N = 3$ and $t = \frac{3}{5}\, \rme^{2\pi\rmi/3}$. All digits displayed are stable. Note that \textit{all} the digits in the perturbative resummation already match the exact solution, so the transseries parameter must be $0$.}
\label{tab:resumjustpert}
\end{table}
Pushing $\arg(t)$ further towards $+\pi$ we cross the second Stokes line. This restores the value of $\sigma$ back to what it was at $\arg(t) = 0$, namely \eqref{valuesigma}. After this point the perturbative result alone is already significantly different from the exact one, as can be clearly seen in figure \ref{fig:resumberryplotFex1}. We could, in principle, stop at $\arg(t) = +\pi$, but we know that the partition function has monodromy $2$ around $t=0$ so we keep pushing the calculation. In order to do this, for the exact value of the free energy we will use the analytic continuation of the hypergeometric function, explained around equation \eqref{confluent_monodromy}. For the transseries we need to do an analytical continuation of both its coefficients and the instanton action. As we move beyond $\arg(t)>+\pi$ the first observation is that perturbation theory becomes less and less accurate. Shortly afterwards we cross an anti-Stokes line, signaling a change in the sign of $\re \left(A(t)/t\right)$, from negative to positive. Now instanton contributions are roughly of the same order of magnitude as the perturbative contribution, and this effect is dramatically enhanced as $\arg(t)$ grows. Following the sequence of points in figure \ref{fig:resumberryplotFex1} we see how each new instanton contribution struggles to move the resummation line closer to the exact curve. As we get to $\arg(t) = +\pi + \frac{\pi}{4}$ not even three instanton terms can give an accurate result, and higher instanton sectors are needed to keep up with the analytical curve.

Figure \ref{fig:resumberryplotFex1} is a very rich picture which involves both Stokes and anti-Stokes lines, regions with $\sigma = 0$ and with $\sigma \neq 0$, and a crisp image of the importance of instanton corrections. In fact it shows how the instantons in the transseries are relevant for much more than achieving exponential accuracy in matching the exact results: they actually need to take \textit{dominance} in order to properly describe the physics at small $N$. This should be extremely compelling evidence towards the relevance of resurgent transseries in describing the nonperturbative realm. We may now proceed with exploring the gauge-theory parameter space. In figures \ref{fig:resumberryplotRex1}, \ref{fig:resumberryplotRex2}, and \ref{fig:resumberryplotRex3}, we expand our list of examples by keeping $N=3$ but varying the value of $|t|$; and this time around we consider the orthogonal polynomial recursion coefficients $\CR(N,t)$. In some of these figures for $\CR(N,t)$ we can actually follow the entire monodromy and ``close'' the curves (the cases we are presently addressing are the ones shown earlier in figure \ref{fig:r3_monodromy}). These figures also show a rather interesting feature: the perturbative resummation has a tendency to follow along the \textit{planar} approximation and thus, past anti-Stokes boundaries, it completely misses the correct features of the problem. It is the instantons that come to save the day and properly describe the monodromy properties we are trying to reproduce. Furthermore, we keep expanding our list of examples in figures \ref{fig:resumberryplotFex3} and \ref{fig:resumberryplotFex2}, by now addressing the free energy at $N=2$ and for new values of $|t|$; and these cases are in correspondence with the exact results shown in figure \ref{fig:F23_monodromy}. In all these figures we can cleanly identify all the Stokes, anti-Stokes, and analytic continuation transitions which occur as we draw the constant $|t|$ paths on the phase diagram of figure \ref{fig:phasediagramv1}.

Finally, we wish to study the partition function as a function of $N$, where $N$ will be taken as an \textit{arbitrary complex number}, and at fixed 't~Hooft coupling $t$. The first thing one can try to do is extend our results towards negative $N$. This is shown in figure \ref{fig:negative_N}, for $t=\frac{1}{2}\, \rme^{\frac{5\pi\rmi}{6}}$. At positive $N$ the different lines smoothly interpolate between the exact results, and it is in fact impossible to tell them apart with the naked eye. Only very close to $N=0$ do we notice the instanton contributions behaving incorrectly. As we have discussed earlier, this is a point of infinitely strong coupling where we do not have enough data---in fact at this point the perturbative resummation is more reliable as we have more than twice the perturbative coefficients as compared to the (multi) instanton coefficients (see the appendix). As one moves towards $N < 0$, we should bear in mind that we have entered a region where $\exp \left( - N A(t)/t \right)$ has changed sign. This means that the perturbative sector is no longer a viable approximation to the full answer, since it just became of the same order as the instanton sectors. The plot in figure \ref{fig:negative_N} illustrates how up until $N \approx -4$ the instanton sectors still look consistent, and are providing corrections which sit on top of the previous sector. However, as $N$ gets increasingly negative we start getting wilder oscillations and the resummation can no longer be trusted. We would need more coefficients, and crucially higher instanton sectors, in order to carry on. In spite of this, it is clear how the transseries resummation allows us to define gauge theory at negative rank!

We can next look at complex values of $N$, and this is shown in figure \ref{fig:complex_N}, for $t=10\, \rme^{\frac{99\pi\rmi}{100}}$ (in this figure we are just plotting the third instanton contribution to the partition function, which visually is in fact indistinguishable from the perturbative or lower instanton resummations). Similarly to what we have found when looking at real $N$, positive or negative, we have a function which oscillates and those oscillations can get milder or harsher as we move away from the real line. Note that in \cite{cgm14} the authors were able to analytically show that the $\CN=8$ ABJ(M) partition function was an entire function of $N$. Crucial to that, as we have mentioned before, was the drastic simplification they found in their nonperturbative structure. The quartic matrix model partition function we address in the present paper has no supersymmetry, milder integrability properties, and a full transseries to deal with, implying that such an analytical proof is at present still unachievable\footnote{Further studies of the partition function of the quartic matrix model will appear in \cite{asv15, csv15}.}. Generic gauge theories will be even more complicated. Nevertheless, we can carry out numerical tests, such as the ones illustrated above, and based on the evidence we have been able to produce we find encouraging signs that the partition function will indeed be an entire function of $N$, at fixed 't~Hooft coupling. Future research should definitely address this issue, perhaps starting by producing more data, both higher $1/N$ coefficients in the sectors we have already computed and in higher instanton sectors. In any case, it is certainly remarkable that our one single transseries was able to correctly reproduce the many different (nonperturbative) monodromy structures at play, for different values of $N$ and $t$. It is also extremely interesting that at the same time it went far beyond any available analytical results by extending the gauge theory to arbitrary complex $N$.

\acknowledgments
We would like to thank Marcos Mari\~no and Marcel Vonk for useful discussions, comments and/or correspondence. The research of RCS and RS was partially supported by the FCT-Portugal grants PTDC/MAT/119689/2010 and EXCL/MAT-GEO/0222/2012. The research of RV was partially supported by the FCT-Portugal grant SFRH/BD/70613/2010.

\newpage

\appendix

\section{Quartic Matrix Model: Transseries Data}\label{sec:appendix}

Our resurgence approach to extract finite $N$ results out of the (nonperturbative) large $N$ expansion was illustrated throughout the main body of the paper within the example of the transseries solution to the one-cut quartic matrix model. In this appendix we briefly overview the data for this gauge theory, and the methods required in order to obtain it. These methods were extensively used in \cite{asv11, sv13} to construct full two-parameter transseries solutions to the one- and two-cut quartic matrix model, resulting in large amounts of data for both perturbative and multi-instanton sectors. For the purposes of our current numerical analysis, however, the full data in \cite{asv11, sv13} is not needed but just about ``half''. The transseries written down in \cite{asv11, sv13} were written as asymptotic expansions in the matrix-model string coupling, $g_s$, being a bit closer in spirit to the string theoretic literature. In this case, remaining closer to the gauge theoretic literature, we shall slightly rewrite these results using the 't~Hooft coupling instead, where $t = g_s N$.

The starting point is the so-called string equation, an equation computing the recursion coefficients $r_n$, \eqref{eq:Zhr}, given some choice of polynomial potential in the matrix integral for the partition function. As discussed earlier, if one is able to compute these coefficients, then the partition function itself follows. For the case of the quartic potential, the string equation takes the form
\be 
r_n \left(1 - \frac{\lambda}{6} \left( r_{n-1} + r_n + r_{n+1} \right) \right) = n g_s.
\ee
\noindent
The standard procedure starts by considering a continuum limit, where we introduce a new continuous variable $x = n g_s$ and where the recursion coefficients $r_n$ get promoted to functions $\CR (x)$ (we shall take the value $x=t$ in the following). This function is then written as a transseries,
\bea
\CR (N,t) &=& \sum_{n=0}^{+\infty} \sigma^{n}\, \CR^{(n)}(N,t), \\
\CR^{(n)} (N,t) &\simeq& \rme^{- n N \frac{A(t)}{t}}\, \sum_{g=0}^{+\infty} N^{-g-\beta_n}\, t^{g+\beta_n} R^{(n)}_g (t),
\eea
\noindent
where $\sigma$ is the transseries parameter and $A(t)$ is the instanton action. The coefficient $\beta_n$ may be regarded as a ``characteristic exponent''. In the present case $\beta_n = n/2$ (but see \cite{asv11} for tables indicating the different values of $\beta_n$ for the transseries of the quartic model). Note that in general one actually needs to consider a \textit{two}-parameter transseries, with instanton actions $\pm A(t)$, leading to generalized instantons as one needs to consider many new nonperturbative sectors $(n) \to (n|m)$. Moreover, all these nonperturbative sectors are related to each other via resurgence; see \cite{asv11, sv13}. For most of the numerical analysis we carry out in this paper, the above one-parameter transseries is enough. Another thing to notice is that the instanton action and the asymptotic coefficients have ``attached'' adequate powers of $t$. This is just so that what we herein call $A(t)$ and $R^{(n)}_g (t)$ are the \textit{exact} same quantities as those computed in \cite{asv11}. Now, by plugging this transseries expansion into the (continuous) string equation\footnote{Not to clutter notation too much, we sometimes omit the first argument in $\CR(N,t)$.}
\be 
\CR (t)  \left\lbrace 1 - \frac{\lambda}{6} \left( \CR \left( t-g_s \right) + \CR (t) + \CR \left( t+g_s \right) \right) \right\rbrace = t,
\ee
\noindent
we can recursively solve for the coefficients $R^{(n)}_g(t)$, as well as compute the instanton action (see \cite{asv11} for details). The equations one finds are differential when $n=1$ and algebraic in all other cases. The coefficients $R^{(n)}_g(t)$ can be written in terms of the variable $r \equiv \frac{1}{\lambda} \left( 1 - \sqrt{1-2 \lambda \, t} \right)$ and they have a pattern of the form
\be 
R^{(n)}_g (t) = \frac{(\lambda\, r )^{p_1}}{r^{p_2} \left( 3-3\lambda\, r \right)^{p_3} \left( 3-\lambda\, r \right)^{p_4}}\, P^{(n)}_g (r).
\ee
\noindent
The exponents in the prefactor are functions of $n$ and $g$,
\be
\label{p_i}
\begin{aligned} 
p_1 &= \frac{1}{2} \left( 3 n - 2 \right), \\
p_2 &= n + g - 1, \\
p_3 &= \frac{1}{4} \left( 5n + 10g - 4 \right), \\
p_4 &= \frac{1}{4} \left( 3n + 6g + 2\delta - 4 \right),
\end{aligned}
\ee
\noindent
with $\delta = n \text{ mod } 2$, and the $P^{(n)}_g(r) $ are polynomials of degree $(n + 6g + \delta - 2)/2$. For the purposes of the numerical analysis in the main text, we have focused only on instanton sectors up to $n=3$. The maximum order of the asymptotic coefficients computed in each sector is shown below, in table \ref{tablequartic}.
\begin{table}[ht]
\centering
\begin{tabular}{c|ccccc}
\begin{picture}(20,15)(0,0)
\put(7,0){$n$}
\end{picture}
& 0 & 1 & 2 & 3 \\
\hline
$g_{\text{max}}$ & 200 & 50 & 50 & 50 
\end{tabular}
\caption{Highest order $g$ for which we have calculated $R^{(n)}_g(t)$.}
\label{tablequartic}
\end{table}
For completeness let us also write down the first coefficients in each sector. For $n=0$,
\be 
R^{(0)}_0 = r, \qquad 
R^{(0)}_2 = \frac{\lambda^2 r}{6 \left(1-\lambda r\right)^4}, \qquad 
R^{(0)}_4 = \frac{7\lambda^4 r \left(5+2\lambda r\right)}{72 \left(1-\lambda r\right)^9}.
\ee
\noindent
With $n=1$ the equation at order $N^{-g}$ gives a solution for $R^{(1)}_{g-1}(t)$. At order $N^0$ we compute the instanton action (see as well \cite{msw07, m08, msw08})
\be 
\label{eq:instantonaction}
A = -\frac{r}{2} \left(2-\lambda r\right) \text{arccosh} \left(\frac{3-2\lambda r}{\lambda r}\right) + \frac{1}{2\lambda} \sqrt{\left(3-\lambda r\right) \left(3-3\lambda r\right)},
\ee
\noindent
and the first coefficients that follow are
\be 
R^{(1)}_0 = \frac{\sqrt{\lambda r}}{\left(3-\lambda r\right)^{1/4} \left(3-3\lambda r\right)^{1/4}}, \qquad 
R^{(1)}_1 = - \frac{9 \sqrt{\lambda r} \left(6-3\lambda r-6\lambda^2 r^2+2\lambda^3 r^3\right)}{8r \left(3-\lambda r\right)^{7/4} \left(3-3\lambda r\right)^{11/4}}.
\ee
\noindent
Finally, for $n=2, 3$, we have
\begin{align} 
& R^{(2)}_0 = -\frac{\lambda^2 r}{2 \left( 3 - \lambda r \right)^{1/4} \left( 3 - 3 \lambda r \right)^{3/2}}, 
& R^{(2)}_1 = \frac{3 \lambda^2 \left( 18 + 117 \lambda r - 102 \lambda^2 r^2 + 22 \lambda^3 r^3 \right)}{8 \left( 3 - \lambda r \right)^2 \left( 3 - 3 \lambda r \right)^4}, \\
& R^{(3)}_0 = \frac{3 \left( \lambda r \right)^{7/2} \left( 2 - \lambda r \right)}{8 r^2 \left( 3 - \lambda r \right)^{7/4} \left( 3 - 3 \lambda r \right)^{11/4}}, 
& R^{(3)}_1 = \frac{27 \left( \lambda r \right)^{7/2} \left( 18 + \lambda r - 12 \lambda^2 r^2 + 4 \lambda^3 r^3 \right)}{16 r^3 \left( 3 - \lambda r \right)^{11/4} \left( 3 - 3 \lambda  r \right)^{19/4}}.
\end{align}

Having addressed the transseries multi-instanton structure of $\CR(N,t)$, one may proceed and address the free energy next. The relation between these quantities is encapsulated in a Toda-like equation (which is in fact nothing more than the continuous version of \eqref{randZ}),
\be
\label{TodaRF}
\CF \left(t+g_s \right) - 2 \CF (t) + \CF \left(t-g_s \right) = \log \left( \frac{\CR(t)}{t} \right).
\ee
\noindent
Here $\CF = F - F_{\text{G}}$ denotes the free energy of the quartic matrix model normalized against the Gaussian contribution. By force of the above relation, the free energy will inherit the transseries structure of $\CR(N, t)$, such that again one finds
\be
\CF (N,t) = \sum_{n=0}^{+\infty} \sigma^{n}\, \CF^{(n)}(N,t),
\ee
\noindent
with
\bea
\CF^{(0)} (N,t) &\simeq& \sum_{g=0}^{+\infty} N^{2-2g}\, t^{2g-2} \CF^{(0)}_{2g} (t), \\
\CF^{(n)} (N,t) &\simeq& \rme^{- n N \frac{A(t)}{t}}\, \sum_{g=0}^{+\infty} N^{-g-\beta_n^{\CF}}\, t^{g+\beta_n^{\CF}} \CF^{(n)}_g (t) \,,
\eea
\noindent
where $\beta_n^{\CF} = n/2 $. Plugging this back in \eqref{TodaRF} and expanding in powers of $\sigma$ we find
\be 
\CF^{(n)} \left( t+g_s \right) - 2 \CF^{(n)} (t) + \CF^{(n)} \left( t-g_s \right) = \Phi^{(n)}(t),
\ee
\noindent
where $\Phi^{(n)}(t)$ is the $n$-th instanton sector of $\log \left(\frac{\CR(t)}{t} \right)$. The first few sectors, which we will be using, are
\be 
\Phi^{(0)}(t) = \log \left(\frac{R^{(0)} (t)}{t} \right), \qquad \Phi^{(1)}(t) =\frac{R^{(1)} (t)}{R^{(0)} (t)}, \qquad \Phi^{(2)}(t) =\frac{R^{(2)} (t)}{R^{(0)} (t)} - \frac{1}{2} \left(\frac{R^{(1)} (t)}{R^{(0)} (t)}\right)^2.
\ee
\noindent
The standard course of action, \textit{e.g.}, \cite{biz80, m08, asv11, sv13}, essentially amounts to inverting \eqref{TodaRF} and finding explicit equations for each transseries component $\CF^{(n)}_g (t)$. Herein we have taken a slightly different route, already starting at the perturbative level, $n=0$, which turns out to be much more computationally efficient at high genus. What we do is simply to use \eqref{TodaRF} directly: start by expanding its right-hand-side in powers of $1/N$, where the expansion of the logarithm is now written as a sum over partitions
\be 
\log \left( \frac{R^{(0)} (t)}{t} \right) = \log \left( \frac{R^{(0)}_0 (t)}{t} \right) + \sum_{k=1}^{+\infty} \frac{t^{2k}}{N^{2k}}\, \sum_{s \geq 1} \frac{(-1)^{s-1}}{s} \sum_{\ell_1 + \cdots + \ell_s = k} \frac{R^{(0)}_{\ell_1} (t)}{R^{(0)}_0 (t)} \cdots \frac{R^{(0)}_{\ell_s} (t)}{R^{(0)}_0 (t)}.
\ee
\noindent
Now on the left-hand-side, at order $1/N^{2g}$, we see that the highest order coefficients cancel out and we are left with a differential equation for $\partial_t^2 \CF^{(0)}_{2g}$. Knowing that the perturbative free energies follow a pattern\footnote{With exceptions at $g=0, 1$.} written in terms of the variable $r$ (see \cite{asv11}),
\be 
\CF^{(0)}_{2g} = \frac{\lambda^{2g-1}}{\left( 2 - \lambda\, r \right)^{2g-2} \left( 1 - \lambda\, r \right)^{5(g-1)}}\, \CP^{(0)}_{2g}(r),
\ee
\noindent
with $\CP^{(0)}_{2g}(r)$ a polynomial of degree $3g-3$, it is a computationally straightforward task to plug this \textit{ansatz} into the equation and solve for the coefficients of the polynomial. The boundary conditions amount to setting $\left. \CF^{(0)}_{2g} \right|_{\lambda=0}=0$, which is a consequence of the Gaussian normalization. The highest order to which we computed $\CF^{(0)}_{2g}$ is shown in table \ref{tableFquartic}. The first coefficients are
\bea
\CF^{(0)}_{0} &=& \frac{r^2}{96} \left( \lambda r \left( 9 \lambda r - 16 \right) + 12 \left( 2 - \lambda r \right)^2 \log \left( \frac{2}{2 - \lambda r} \right) \right), \\
\CF^{(0)}_{2} &=& -\frac{1}{12} \log \left(\frac{2-2 \lambda r}{2-\lambda r}\right), \\
\CF^{(0)}_{4} &=& \frac{\lambda^3 \left( 41 \lambda^2 r^3 - 185 \lambda r^2 + 200 r \right)}{2880 \left( \lambda r-2 \right)^2 \left( 1-\lambda r \right)^5}.
\eea
\noindent
Proceeding with the one and two instanton coefficients, the calculation follows in a straighforward fashion. The equations at order $\sigma^n\, g_s^g$ are now algebraic equations for $\CF^{(n)}_{g}$, and we solve them using the same method of plugging-in an \textit{ansatz} and reducing the problem to one of finding coefficients of a polynomial. The maximum orders of the coefficients we found for $n=1, 2, 3$ are shown in table \ref{tableFquartic}. Essentially, we are limited by the number of $R^{(n)}_g$ we calculated beforehand. 
\begin{table}[h]
\centering
\begin{tabular}{c|ccccc}
\begin{picture}(20,15)(0,0)
\put(7,0){$n$}
\end{picture}
& 0 & 1 & 2 & 3\\
\hline
$g_{\text{max}}$ & 130 & 50 & 50 & 50 
\end{tabular}
\caption{Highest order $g$ for which we have calculated $\CF^{(n)}_g(t)$.} 
\label{tableFquartic}
\end{table}
The functions $\CF^{(n)}_{g}$, which we computed for $n=1,2,3$, are of the form\footnote{We note that there is a small typo in the formulae for the free energy coefficients in \cite{asv11}. The factors of $t - \alpha^2$ therein, where $\alpha^2=r$, should in fact be $\alpha^2 - t$.}
\be 
\CF^{(n)}_{g} = \frac{(\lambda\, r)^{p_1+1}}{r^{p_2+1} \left( 3 - 3\lambda\, r \right)^{p_3+1} \left( 3 - \lambda\, r \right)^{p_4+1-\delta}}\, \CP^{(n)}_{g}(r), \qquad n \geq 1,
\ee
\noindent
where the $\CP^{(n)}_{g}(r)$ are polynomials of degree $( 6g+n-\delta )/2$, and the exponents $p_i$ were defined in \eqref{p_i}. At lowest order, the $n=1, 2$ and $3$ coefficients are 
\bea 
\CF^{(1)}_{0} &=& \frac{\lambda^{3/2}\, r^{1/2}}{2 \left( 3-3\lambda  r \right)^{5/4} \left( 3-\lambda r \right)^{1/4}}, \\
\CF^{(1)}_{1} &=& - \frac{9 \lambda^{3/2} \left( 6 + 75 \lambda r - 54 \lambda^2 r^2 + 10 \lambda^3 r^3 \right)}{16 r^{1/2} \left( 3 - 3\lambda r \right)^{15/4} \left( 3 - \lambda r \right)^{7/4}}, \\
\CF^{(2)}_{0} &=& - \frac{\lambda^3 r \left( 3 - 2 \lambda r \right)}{8 \left( 3 - 3\lambda r \right)^{5/2} \left( 3 - \lambda r \right)^{3/2}}, \\
\CF^{(2)}_{1} &=&  \frac{3\lambda^3 \left( 54 + 531 \lambda r - 846 \lambda^2 r^2 + 462 \lambda^3 r^3 - 92 \lambda^4 r^4 \right)}{32 \left( 3 - 3\lambda r \right)^5 \left( 3 - \lambda r \right)^3}, \\
\CF^{(3)}_{0} &=& -\frac{\lambda^{9/2} r^{3/2} \left(6 - 5 \lambda r\right)}{48 \left( 3-3\lambda r \right)^{15/4} \left(3-\lambda r\right)^{7/4}}, \\
\CF^{(3)}_{1} &=& \frac{3\lambda^{9/2} r^{1/2} \left(-108-1044 \lambda r+1917 \lambda^2 r^2-1086 \lambda^3 r^3+212 \lambda^4 r^4\right)}{128 \left(3-3\lambda r\right)^{25/4} \left(3-\lambda r\right)^{13/4}}.
\eea

Finally, let us note that even though we have presented several results for the partition function in the main body of the paper, we have actually not addressed its transseries representation. In fact, it is computationally very inefficient to exponentiate the free energy transseries and then extract the transseries coefficients for the partition function. If we had done that, we would have ended up only going to much lower orders than what we have achieved for the free energy. As such, we have instead always performed any required resummation first, for the free energy transseries, and only then exponentiated the result.

\newpage

\bibliographystyle{plain}

\end{document}